\documentclass[twocolumn,twocolappendix]{aastex63}
\usepackage{amsmath}

\newcommand\teff{$T_{\rm eff}$}
\newcommand\logg{$\log g$}

\newcommand\vsini{$v\sin i$}
\newcommand\rsini{${\rm R}\sin i$}
\newcommand\sini{$\sin i$}

\newcommand\msun{M$_\odot$}
\newcommand\rsun{R$_\odot$}

\defcitealias{kounkel2022a}{Paper I}
\usepackage{listings}

\begin{document}
\title{Measurement of the angular momenta of pre--main-sequence stars: early evolution of slow and fast rotators and empirical constraints on spin-down torque mechanisms}

\author[0000-0002-5365-1267]{Marina Kounkel}
\affil{Department of Physics and Astronomy, Vanderbilt University, VU Station 1807, Nashville, TN 37235, USA}
\email{marina.kounkel@vanderbilt.edu}
\author[0000-0002-3481-9052]{Keivan G.\ Stassun}
\affil{Department of Physics and Astronomy, Vanderbilt University, VU Station 1807, Nashville, TN 37235, USA}
\author{Lynne A. Hillenbrand}
\affiliation{Department of Astronomy, California Institute of Technology, Pasadena, CA 91125, USA}
\author[0000-0001-9797-5661]{Jes\'us Hern\'andez}
\affiliation{Universidad Nacional Aut\'onoma de M\'exico, Instituto de Astronom\'ia, AP 106,  Ensenada 22800, BC, México}
\author[0000-0001-7351-6540]{Javier Serna}
\affiliation{Universidad Nacional Aut\'onoma de M\'exico, Instituto de Astronom\'ia, AP 106,  Ensenada 22800, BC, México}
\newcommand{\columbia}{Department of Astronomy, Columbia University, 550 West 120th Street, New York, NY 10027, USA}
\author[0000-0002-2792-134X]{Jason Lee Curtis}
\affiliation{\columbia}

\begin{abstract}
We use TESS full-frame imaging data to investigate the angular momentum evolution of young stars in Orion Complex. We confirm recent findings that stars with rotation periods faster than 2~d are overwhelmingly binaries, with typical separations of tens of AU; such binaries quickly clear their disks, leading to a tendency for rapid rotators to be diskless. Among (nominally single) stars with rotation periods slower than 2~d, we observe the familiar, gyrochronological horseshoe-shaped relationship of rotation period versus $T_{\rm eff}$, indicating that the processes which govern the universal evolution of stellar rotation on Gyr timescales are already in place within the first few Myr. Using spectroscopic $v\sin i$ we determine the distribution of $\sin i$, revealing that the youngest stars are biased toward more pole-on orientations, which may be responsible for the systematics between stellar mass and age observed in star-forming regions. We are also able for the first time to make empirical, quantitative measurements of angular momenta and their time derivative as functions of stellar mass and age, finding these relationships to be much simpler and monotonic as compared to the complex relationships involving rotation period alone; evidently, the relationship between rotation period and $T_{\rm eff}$ is largely a reflection of mass-dependent stellar structure and not of angular momentum per se. Our measurements show that the stars experience spin-down torques in the range $\sim10^{37}$~erg at $\sim$1~Myr to $\sim10^{35}$~erg at $\sim$10~Myr, which provide a crucial empirical touchstone for theoretical mechanisms of angular momentum loss in young stars.
\end{abstract}


\section{Introduction}

Young stars are born from large clouds of gas, and even at the earliest stages, these pre-stellar objects exhibit rotation \citep{covey2005}. As the natal core transfers material through the protoplanetary disk onto the protostar sitting at the center, and the protostar itself contracts, through the conservation of angular momentum ($L$), the rotational speed would have increased until reaching break-up velocity. However, there are several mechanisms in place to regulate its rotation. A protostar is tidally locked to the inner part of the protoplanetary disk, matching its Keplerian rotation at the inner wall, removing the excess angular momentum through winds and jets that form bipolar outflows \citep{bouvier1997,matt2012,gallet2019}.

After the outer envelope is consumed, and a young star eventually depletes its disk, and, in the absence of another reservoir that would enable mass to flow, the angular momentum would not increase further. This young star will continue to contract, reducing its size by a factor of few until it settles onto the main sequence. Despite this, the vast majority of stars will not rotate more rapidly over this time as a result of the contraction, as their angular momentum is decreasing due to magnetic braking \citep{skumanich1972}. In large part, this spin down of the angular momentum is very regular as a function of mass, enabling to estimate the age of a star through the measurement of its rotational period via gyrochronology \citep[e.g.,][]{barnes2003}.

There have been many subsequent efforts over the past 20 years to constrain gyrochronology relations, both theoretically \citep[e.g.,][]{gossage2021}, and empirically \citep[e.g.,][]{angus2020,curtis2020}, but they mainly focused on main sequence stars, with only a limited consideration for pre-main sequence stars. Recently, \citet[hereafter, Paper I]{kounkel2022a} have used TESS full frame images (FFI) to measure rotational periods of $\sim$100,000 stars with known ages, producing the largest such catalog to-date that enabled the analysis angular momentum evolution of stars between ages of $\sim$10 Myr to $\sim$1 Gyr. Stars younger than that present a particular complexity due to their rapid evolution, requiring a more careful analysis of their ages, as such they were largely ignored in the analysis.

In this paper, we seek to fill in this gap in the current understanding of the rotational evolution of young stars. In particular, we examine rotation of stars within the Orion Complex, which is one of the closest massive star forming regions, spanning the area on the sky between $76^\circ<\alpha<88^\circ$ and $-10^\circ<\delta<15^\circ$ containing multiple sub-populations with ages from 1 to 10 Myr \citep{kounkel2018a}. It covers the crucial age that is often missed by previous gyrochronology studies. These data enable a precise measurement of a number of parameters beneficial for characterization of stellar rotation in this region.

Stellar rotation in Orion has been examined repeatedly in the past \citep[e.g.,][]{stassun1999,rebull2001,herbst2002}, although the completeness of the census and the level of precision of the underlying data cannot be compared to the modern surveys. More recently, \citet{serna2021} have also revisited rotation in this region. They have measured rotational periods for 517 member of the Orion Complex using TESS FFI, of which 352 stars also had available rotational velocity \vsini\ measurements. As expected, they find a significant anti-correlation between rotation period and \vsini. They also note that in the stars younger than $<5$ Myr, \vsini\ appears to systematically decrease, through some type of angular momentum extraction mechanism (such as through disk locking), but that at the ages older than that, \vsini\ increases due to conservation of angular momentum. Additionally, accreting/disk-bearing young stars (Classical T Tauri stars - CTTSs / Class II) appear to rotate slower than non-accreting/diskless stars (Weak-lined T Tauri stars - WTTSs / Class III) during the first 1.5 Myr, but they appear to have a more comparable distribution at later ages.

While \citet{serna2021} have primarily examined the distribution of \vsini, in this work we aim to examine the rotational period evolution of these young stars, significantly expanding the sample. We also leverage empirically determined stellar radii from \citet{kounkel2020} and precise individual age estimates newly determined by \citet{mcbride2021} to quantify the stellar angular momenta and its evolution as a function of stellar mass and age. 

In Section~\ref{sec:data} we summarize the sources of photometric and spectroscopic data we use to assemble our study sample of $\sim$9000 stars with rotation periods, precise individual ages between 1--10~Myr, disk classifications, and empirical radius determinations. Section~\ref{sec:results} presents the principal results of our investigation, and in Section~\ref{sec:disc} we discuss those results in the context of empirical constraints on theoretical spin-down torque mechanisms, understanding the nature of the rapid rotators, the relationship of rotation to stellar disks, and an intriguing bias in the distribution of stellar inclination angles as a function of age. We conclude with a summary of our findings in Section~\ref{sec:conclusion}. 

\section{Data}\label{sec:data}

\begin{figure*}
\epsscale{1.15}
\plottwo{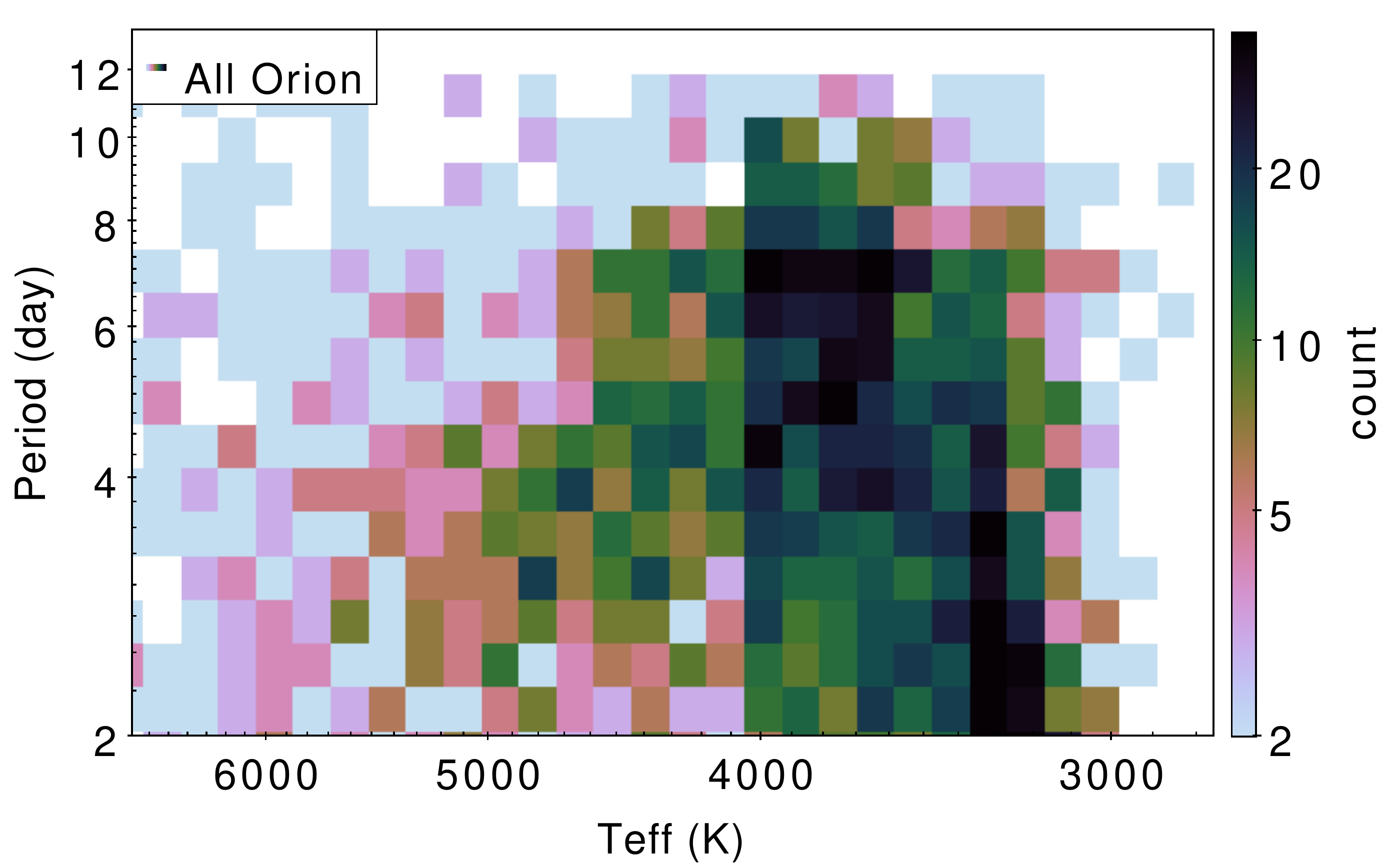}{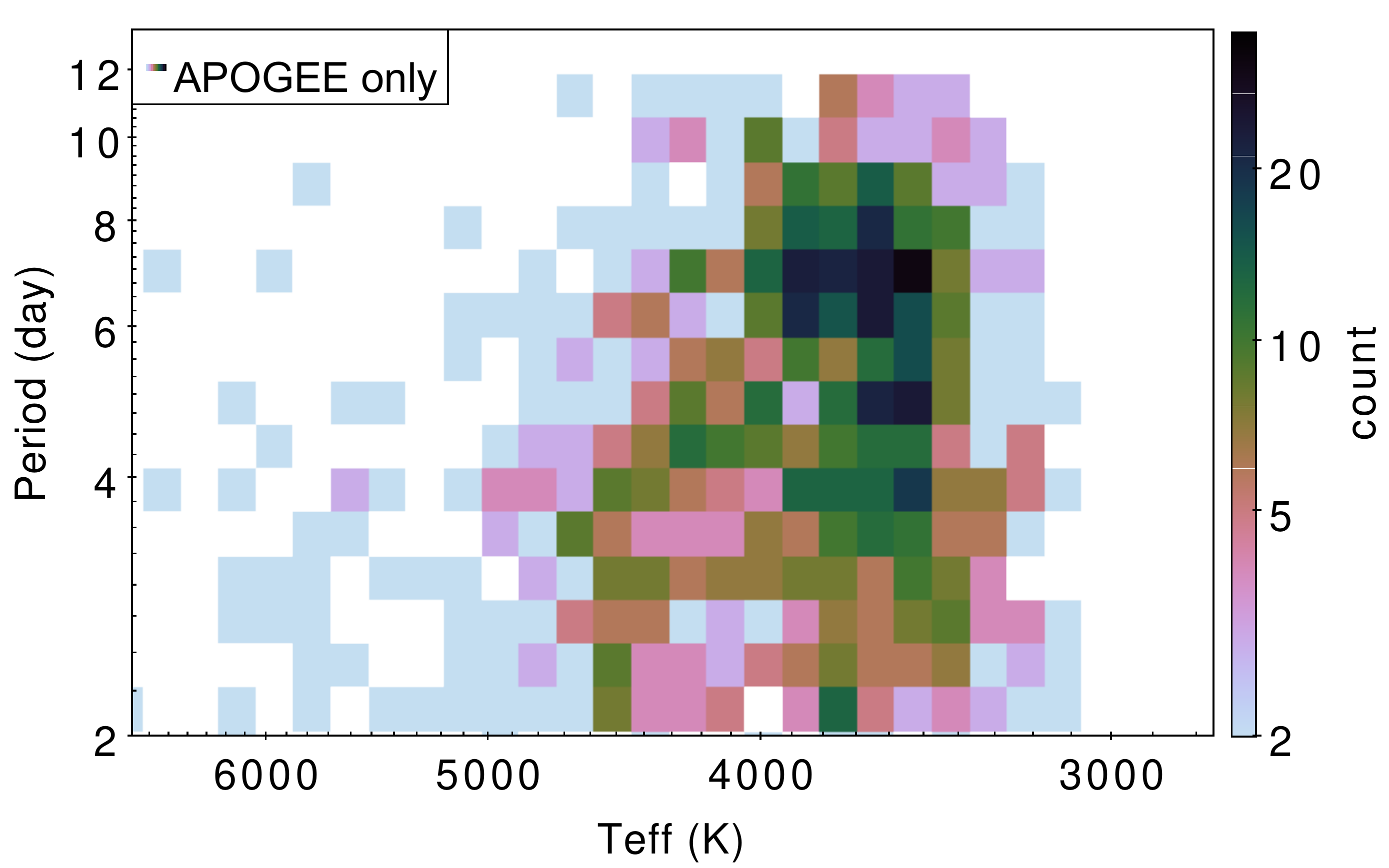}
\caption{Distribution of rotational periods in the sample as a function of \teff. Left panel: all sources in the Orion Complex from \citepalias{kounkel2022a}, using \teff\ estimate from TIC \citep{stassun2019}. Right: a subset of sources observed by APOGEE, with spectroscopically determined \teff.
\label{fig:grid}}
\end{figure*}

The latest census of kinematically selected members of the Orion Complex consists of over 10,000 optically bright stars \citep{kounkel2020}. \citetalias{kounkel2022a} have analyzed TESS FFI data for all of these sources, and have observed periodic signatures in the light curves for $\sim$5700 stars of this sample. As these stars are young, they tend to have strong magnetic fields resulting in large spots, resulting in obvious rotational signatures. Furthermore, with typical rotational periods $<$10 days for the stars at these ages, full period data can easily be recovered even in a single TESS sector. The distribution of rotational periods for the Orion Complex \citepalias{kounkel2022a} is shown in Figure~\ref{fig:grid}.  

APOGEE spectra are available for several thousand stars across the Complex \citep{da-rio2016,cottle2018,kounkel2018a}. APOGEE is a high resolution ($R\sim$22,500) spectrograph covering H-band, operating as part of Sloan Digital Sky Survey \citep{wilson2019}. These data have been extensively analyzed, improving the fidelity of spectral parameters for these young stars, in comparison to the primary pipeline used in public releases \citep{abdurrouf2022}. APOGEE Net \citep{olney2020,sprague2022} provides \teff\ and \logg\ estimates that have been calibrated relative to the pre-main sequence isochrones. \citet{kounkel2019} have estimated rotational velocity \vsini\ of the stars. They also characterized multiplicity within this sample, including identification of both single-lined (SB1) and double-lined (SB2) spectroscopic binaries. In total, APOGEE spectra of 1831 periodic variables from \citetalias{kounkel2022a} in the Orion Complex are available. We adopt these 1831 stars as the primary sample throughout the paper, as only they have the full set of necessary measurements described in this section.

In addition to the directly measured spectroscopic parameters, supplementary data for these stars have been computed. In particular, \citet{kounkel2018a} have estimated angular diameters for these stars through the fitting of the spectral energy distribution (SED), using an atmospheric model corresponding to \teff\ of each star. Coupled with trigonometric parallaxes, this results in a direct estimate of stellar radii. This is an independent radius measurement to those also provided as part of Gaia DR3 \citep{fouesneau2022} or in the TESS Input Catalog \citep{stassun2019}. While all three have comparable estimates, the radii in \citet{kounkel2018a} can be considered more robust for this sample, as a) they are obtained through spectroscopically derived \teff\ to identify the best model \citep[in field stars, this can produce $<$2\% precision;][]{stassun2017}, b) the SED fit spans all of the available fluxes across the full electromagnetic spectrum, c) a special consideration is given to the stars with IR excess, which would affect the SED fit.

However, \teff\ used for the SED fitting in \citet{kounkel2018a} were derived using the IN-SYNC pipeline \citep{cottaar2014}, which had a number of known systematics. \teff\ from the APOGEE Net pipeline are more reliable for these young stars, they offer a better calibration relative to not only the isochrones but also to the more evolved field star. APOGEE Net \teff\ are on average $\sim$300 K cooler than the IN-SYNC \teff. As such, we renormalize the previously reported radii using updated \teff through Stefan–Boltzmann law, resulting in $R$ estimate that are on average 20\% larger than previously reported. We note that the effect of spots is not considered; we discuss the implications from them in Section \ref{sec:inclination}.

We identify all of the known disk-bearing and/or accreting stars in the sample using a combination of different catalogs. They include classification based on H$\alpha$ emission \citep{suarez2017,briceno2019}, classification of NIR excess from Spitzer and other high sensitivity surveys of various regions  \citep{hernandez2007,hernandez2010,megeath2012,grosschedl2019}, as well as a classification of NIR excess based on WISE data across the entire sky \citep{marton2016}. While accretion signature and disk classification are not necessarily interchangeable, they are expected to be equivalent in a vast majority of cases. Together, these correspond to 580 out of 1831 stars in our sample, which we consider to be Class II/CTTS. The remaining sources are considered to be Class III/WTTS. 

We estimate masses for this sample using MassAge (Hernandez J. et al. in prep), through comparing \teff\ and bolometric luminosity to MIST isochrones \citep{choi2016}. Finally, we estimate ages of stars using neural net Sagitta \citep{mcbride2021}. This neural net is utilizing Gaia \& 2MASS photometry (G, BP, RP, J, H, K), Gaia parallaxes, and the typical extinction along the line of sight to predict the likelihood of a star being pre-main sequence, as well as the age of a young star, up to $\sim$80 Myr. It was trained on the census of pre-main sequence stars found in moving groups from \citet{kounkel2020a}, with the age of the stars typically inherited from their parent population (subdividing the moving groups into smaller meaningful populations with a narrow age spread), supplemented with the sources ages of which have been determined independently in literature. Appendix \ref{sec:age} motivates the selection and compares this age model to other techniques. As masses and ages are derived using different models, this may introduce some systematics. But, given that these stars are typically moving vertically down the Hayashi tracks without a significant change in \teff, there should not be a significant variance with mass as a function of age.

\begin{figure*}
\epsscale{1.15}
\plottwo{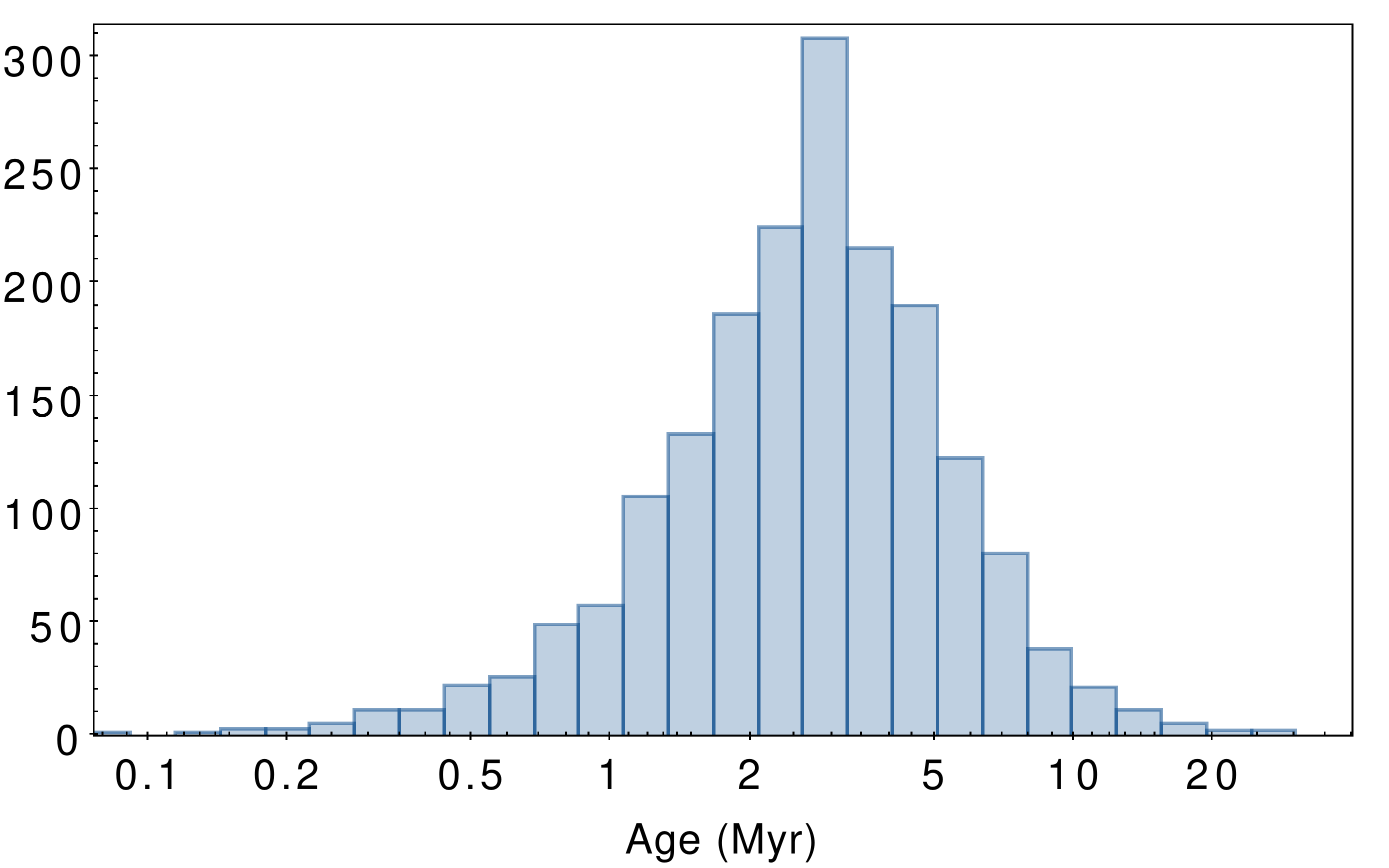}{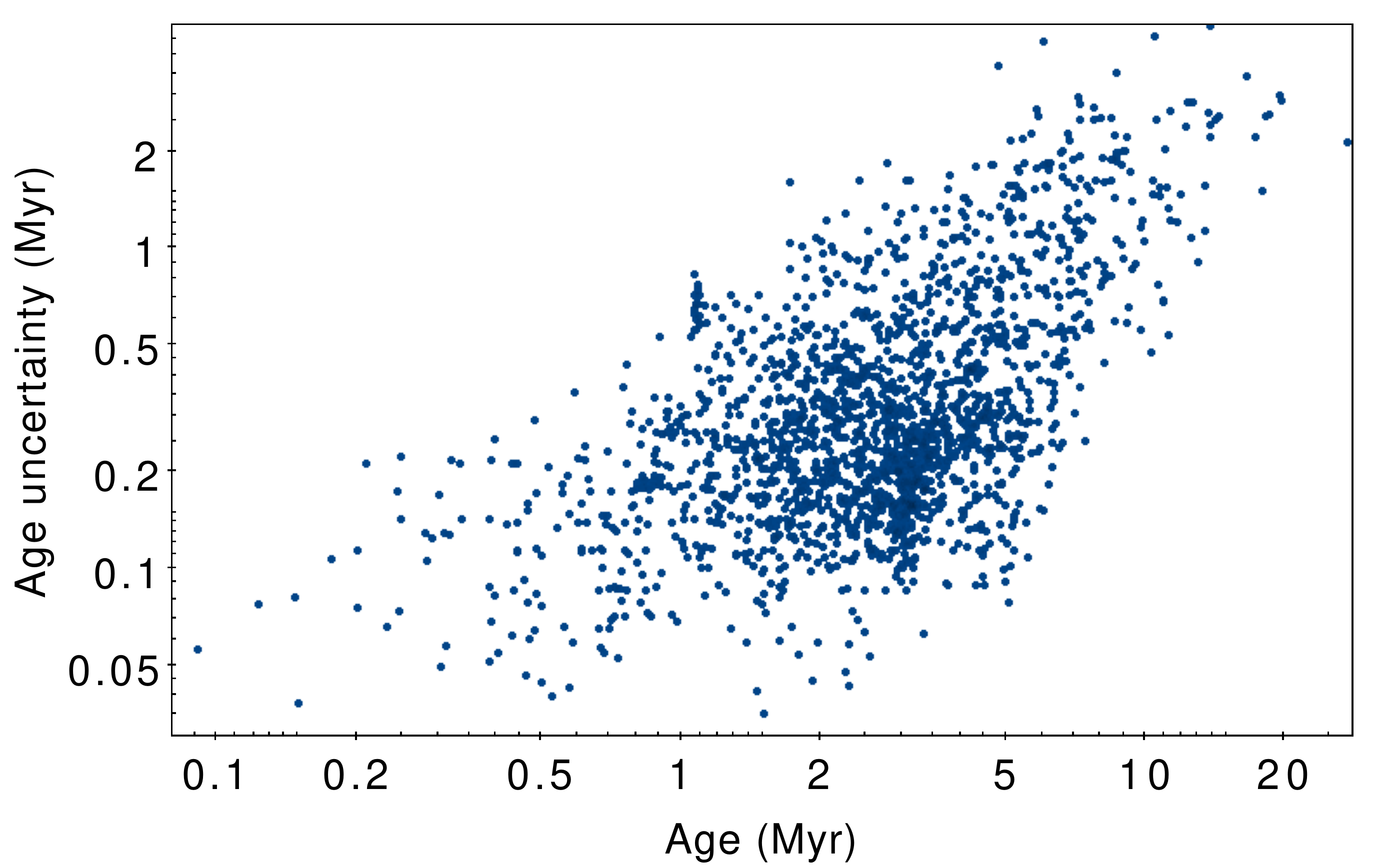}
\caption{Left: Distribution of the adopted ages for the stars in Orion Complex. Right: Typical uncertainties in the sample.
\label{fig:agehist}}
\end{figure*}

The distribution of these ages for the stars in Orion is shown in Figure \ref{fig:agehist}. The uncertainties are generated through propagating the typical uncertainties in the input astrometry and photometry through the model, they typically have a range of 0.05-07 dex. We do note that there is some overlap between the stars that have originally been used for training Sagitta and the stars in this work; we do not expect this to significantly affect the quality of the derived ages.

\section{Results}\label{sec:results}

In this section, we present the results of the distributions of stellar rotation periods for our study sample of Orion Complex stars, typically having ages $\lesssim$10~Myr. The sample includes stars that are rapid rotators and slow rotators, stars that have disks (CTTSs) and those that lack disks (WTTSs), and stars that are single or in binaries. We consider these subsets and their relationships in turn. 

However, to begin, we note that the overall distribution of stellar rotation periods among the nominally single stars in the sample (Figure~\ref{fig:grid}) already at these very young ages manifest the familiar ``horseshoe-shaped" pattern in the $P_{\rm rot}$ vs.\ $T_{\rm eff}$ diagram that has become the basis for gyrochronology at later ages \citep[e.g.,][]{barnes2003}. This suggests that, despite significant scatter at any given age, the dominant physical mechanisms that govern the evolution of angular momentum are already in place and very quickly sculpting the distribution of angular momenta into the universal patterns with stellar mass and age that make gyrochronology possible. 

\subsection{Rapid rotators}

\begin{figure}
\epsscale{1.15}
\plotone{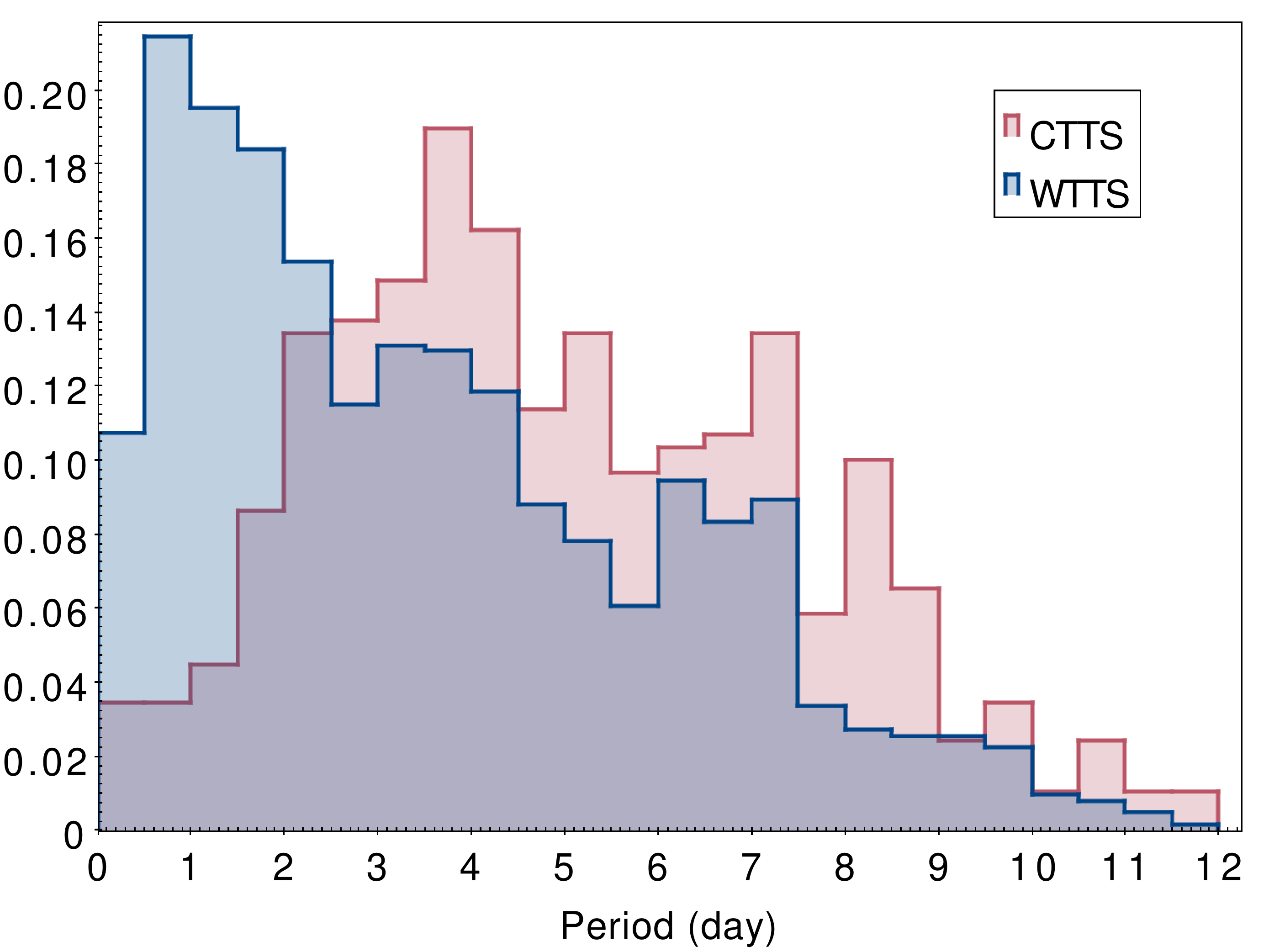}
\caption{Period distribution between CTTS and WTTS objects.
\label{fig:period}}
\end{figure}

Examining the period distribution of CTTS and WTTS objects in Figure \ref{fig:period}, there is a clear difference between the two samples. Very few CTTS have rotational periods faster than 2 days. On the other hand, $\sim$1/3 of WTTSs have rotational periods $<$2 days. 
We use this 2 day period to delineate between ``slow'' and ``fast'' rotators. 

\begin{figure}
\epsscale{1.15}
\plotone{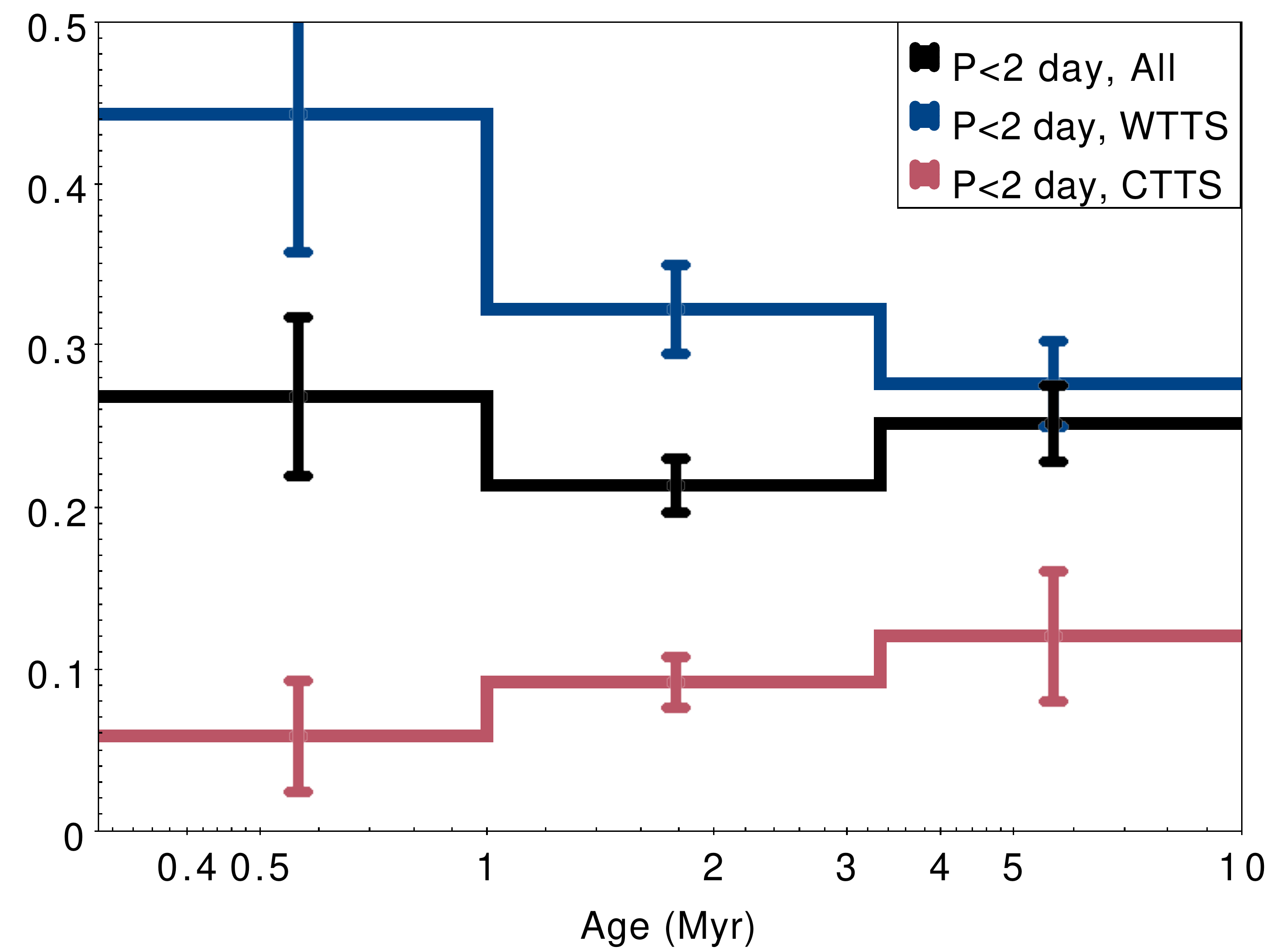}
\plotone{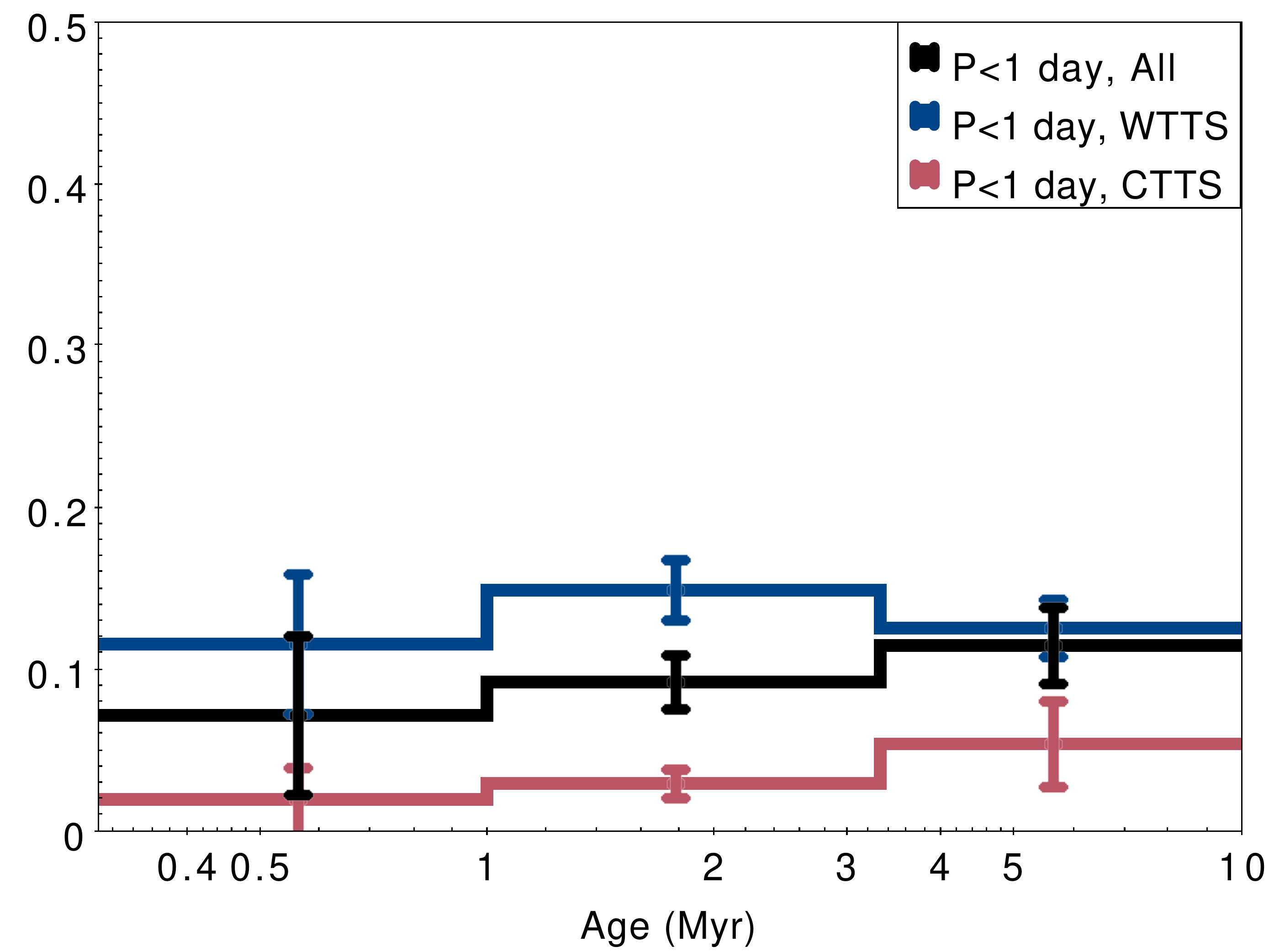}
\caption{Fraction of rapid rotators among CTTSs and WTTSs with \teff$<$6700 K as a function of age. Top panel shows all rapid rotators with periods $<2$ days, bottom panel is restricted to the subset with periods $<1$ day.
\label{fig:rapid}}
\end{figure}

Examining their distribution as a function of age (Figure~\ref{fig:rapid}), we find that the overall fraction of rapid rotators remains more or less constant, however there are important differences between the WTTSs and CTTSs with regards to this trend.
Among WTTSs, the overall fraction of fast rotators decreases from $\sim$45\% at ages $<$1 Myr to $\sim$30\% at ages $>3$ Myr. 
The reason for this trend is likely to be increasing dilution of the diskless stars over time by more slow rotators, as all stars irrespective of their rotation eventually deplete their disks,
until the age where almost all stars are diskless \citep[$\sim$3~Myr; see, e.g.,][]{ribas2014}.
Among CTTSs, there is an opposite trend: there are almost no rapid rotators with disks at an age $<$1 Myr, but their fraction increases moderately to $\sim$10\% at ages $>3$ Myr. 
We return to discuss the implications of this trend in Section~\ref{sec:raprot}. 

At ages $<$1 Myr, rapid rotators tend to have rotational periods between 1 and 2 days. As they age, however, the fraction of systems with periods $<1$ day increase relative to the total number of stars. This process appears to continue at older ages, as in populations older than $\sim$20--30 Myr, most rapid rotators appear to favor periods $<$0.5 days \citepalias{kounkel2022a}. Most likely, without an effective method of reducing a sufficient fraction of their angular momentum, the stars continue to spin up as they shrink towards their main sequence radius, and their spin-up increases until the magnetic braking can overcome this speed increase, or until they stop contracting \citep[e.g.,][]{rebull2022}.

\subsection{Slow rotators}

\begin{figure*}
\epsscale{1.}
\plottwo{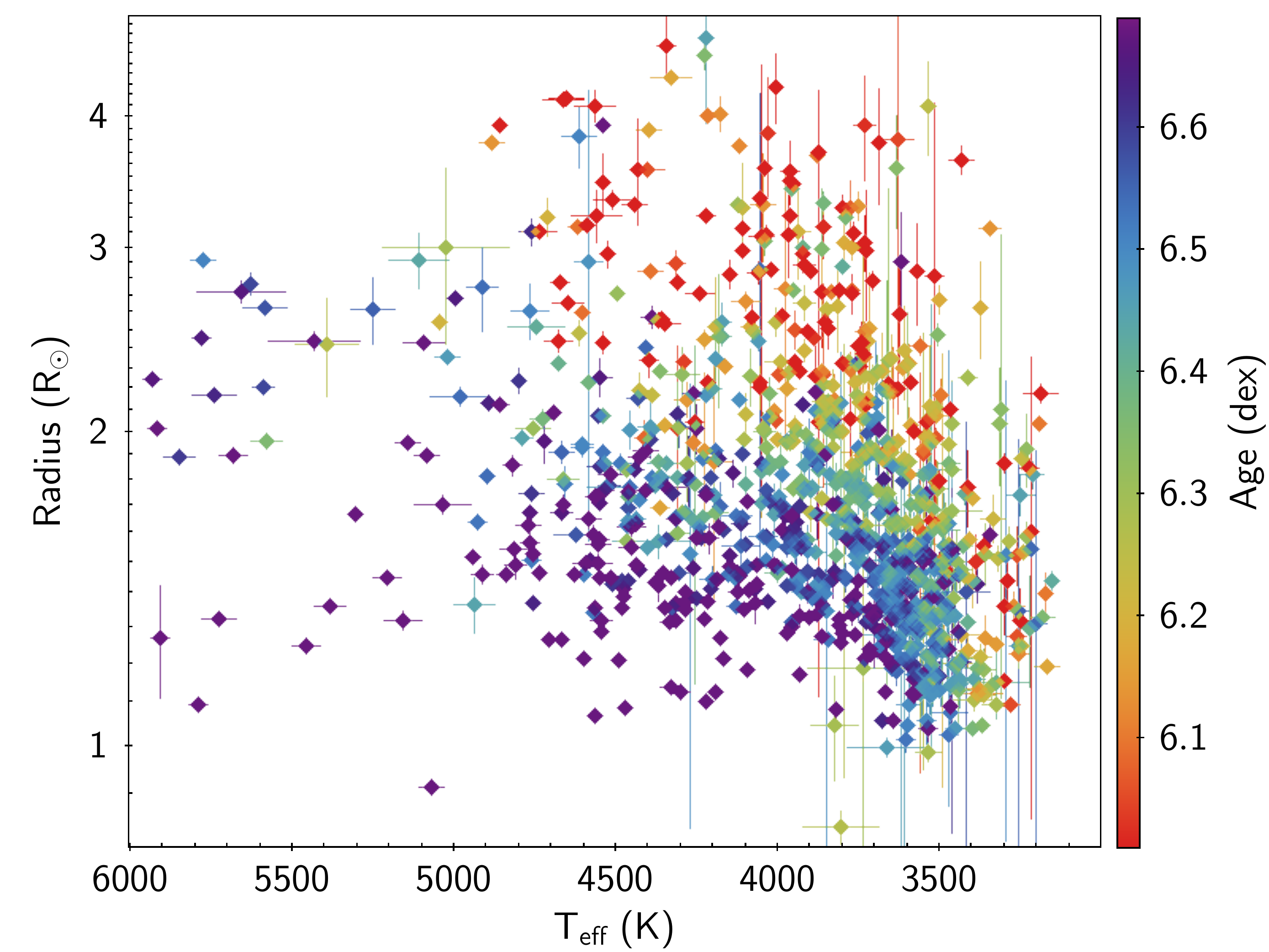}{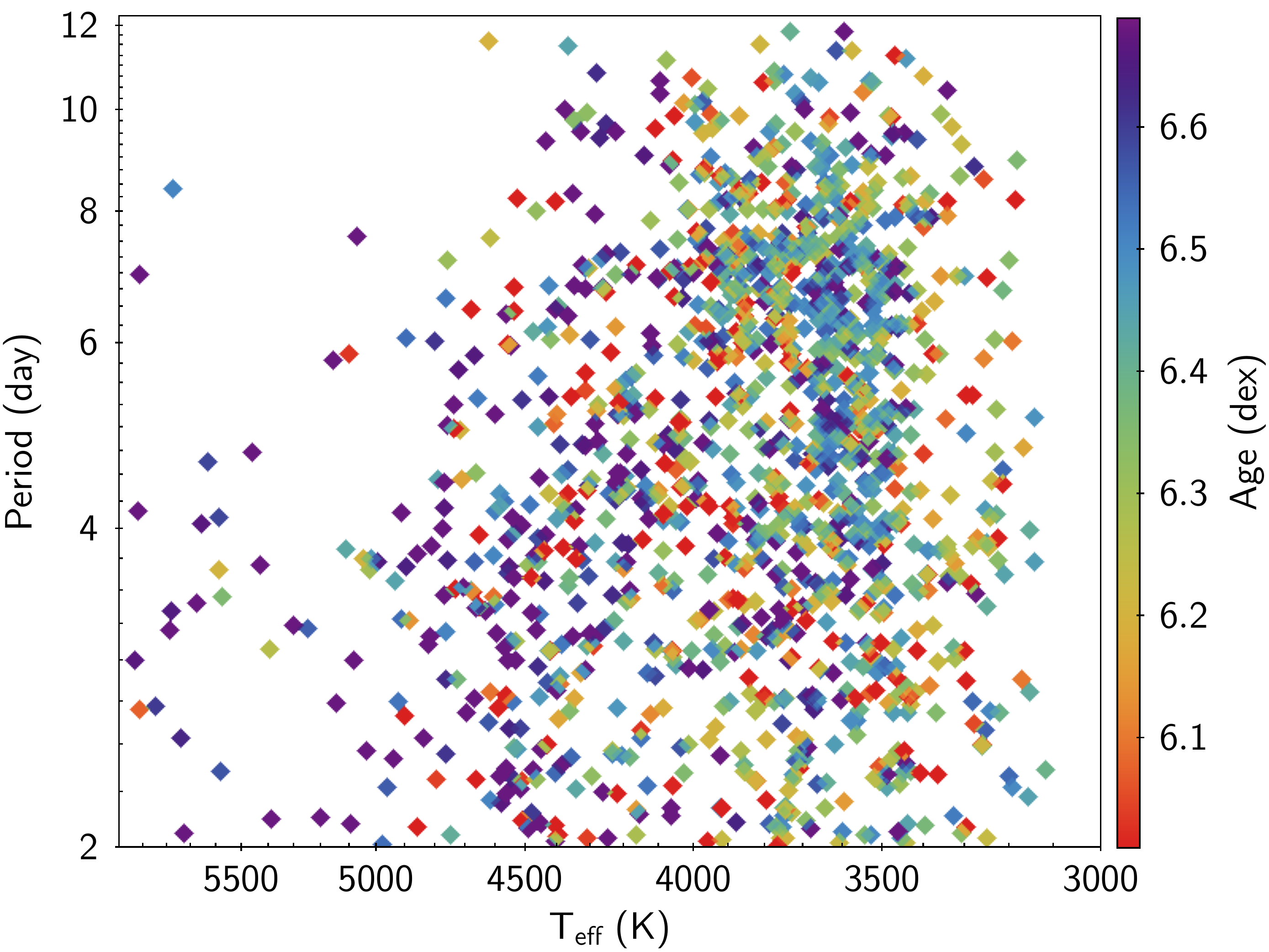}
\plottwo{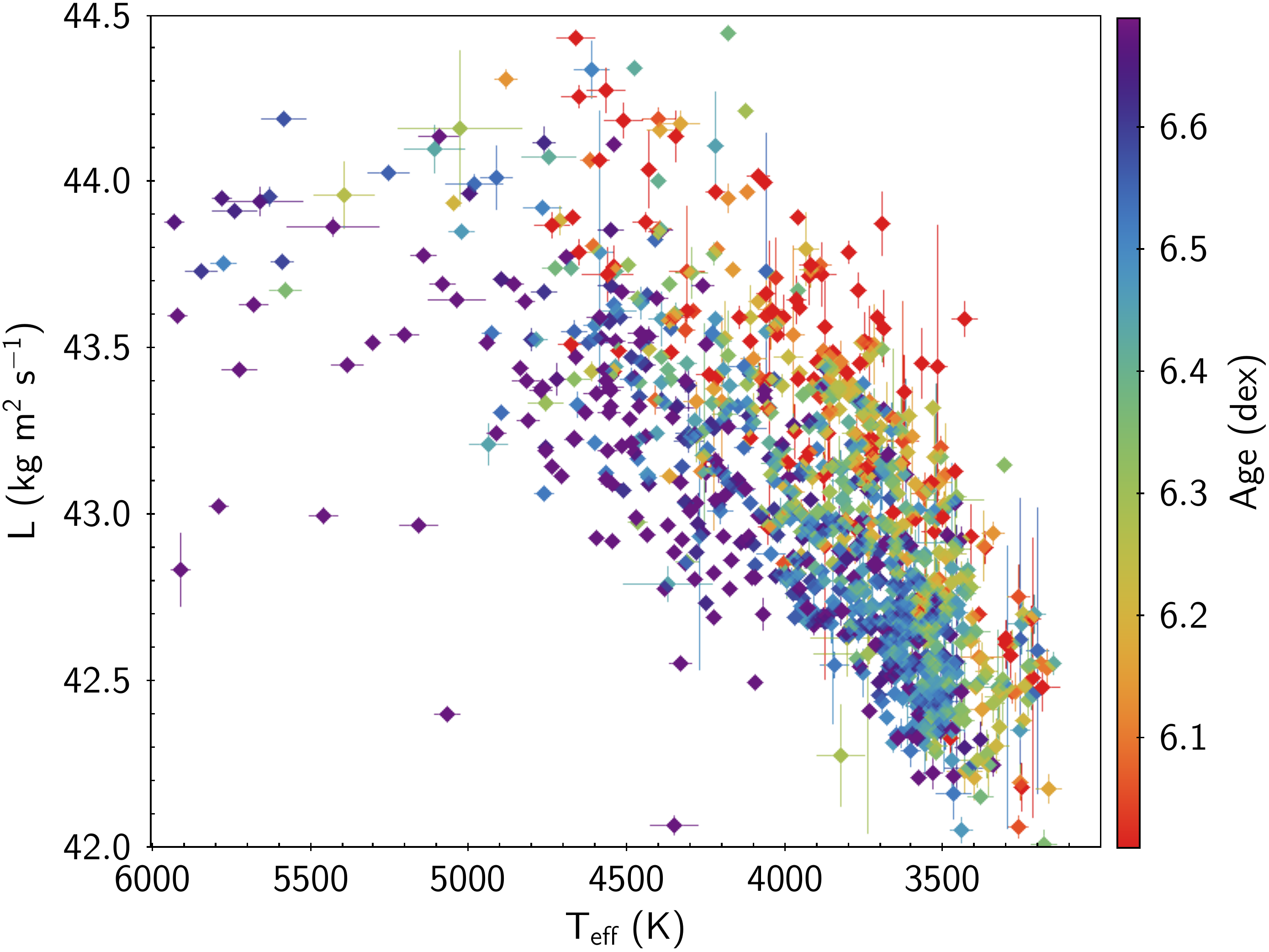}{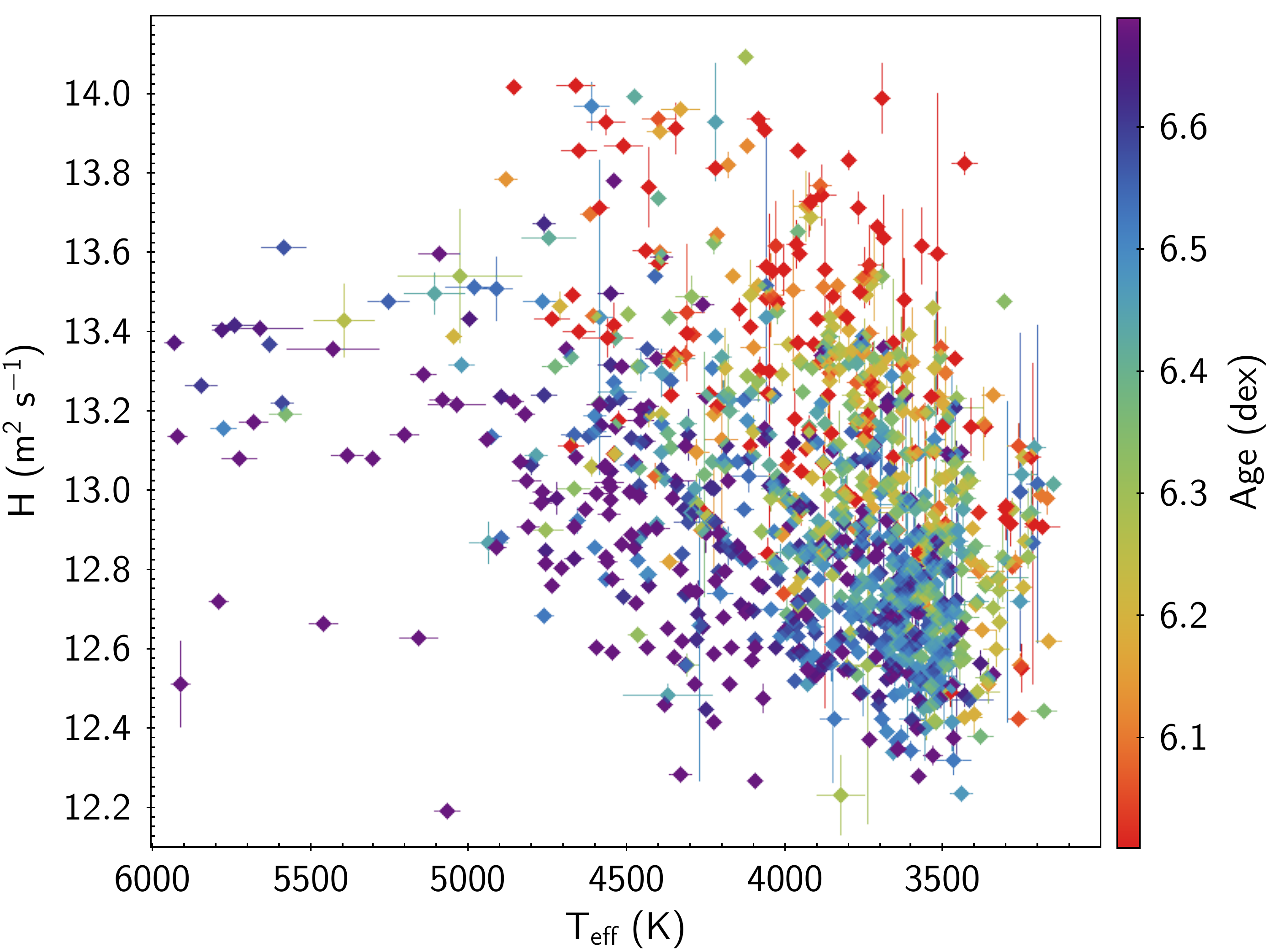}
\caption{Dependence of radius, rotational period, angular momentum, and specific angular momentum on temperature and age of stars in Orion with $P>2$ days. The typical uncertainties in the data are 30 K in \teff, 0.1 \rsun\ in radius, 0.01 day in period, 0.05 dex in L, and 0.04 dex in H. In young disk-bearing stars typical uncertainties in radius are larger due to the infrared excess affecting the quality of the SED fitting.
\label{fig:orion_sangmom}}
\end{figure*}

\begin{figure*}
\epsscale{1.15}
\plotone{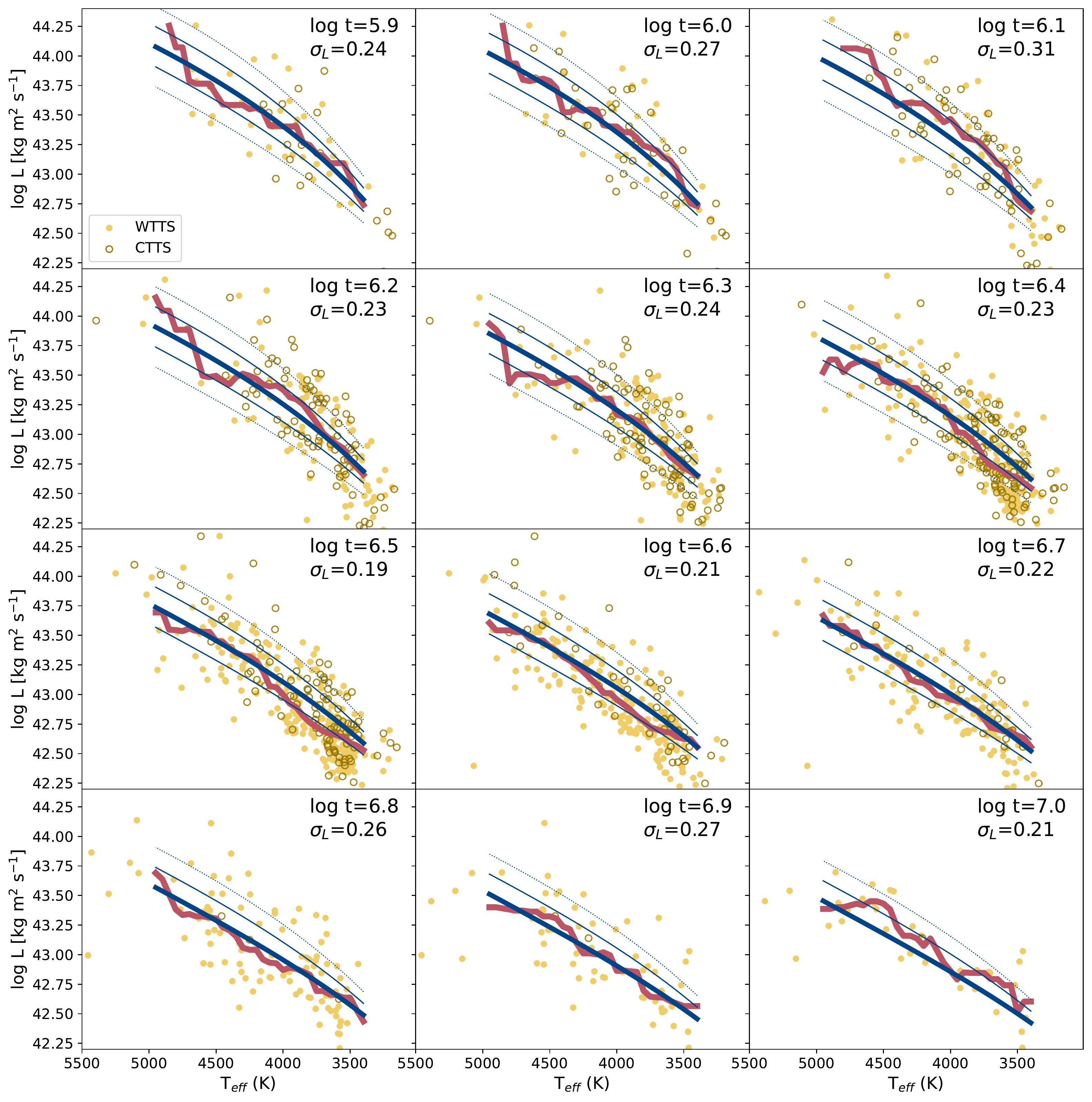}
\caption{A fit of angular momentum L. Yellow dots correspond to the data within each age bin. The red line shows the running median within the data. Thick blue line shows the best fit for L for a given age. Thin blue lines are offset in age by $\pm$0.3 dex, and dotted lines are offset in age by $\pm$0.6 dex. The typical scatter between the data and the resulting fit of L is shown in each panel. Open circles represent CTTSs, filed circles represent WTTSs.
\label{fig:interp}}
\end{figure*}

\begin{figure}
\epsscale{1.15}
\plotone{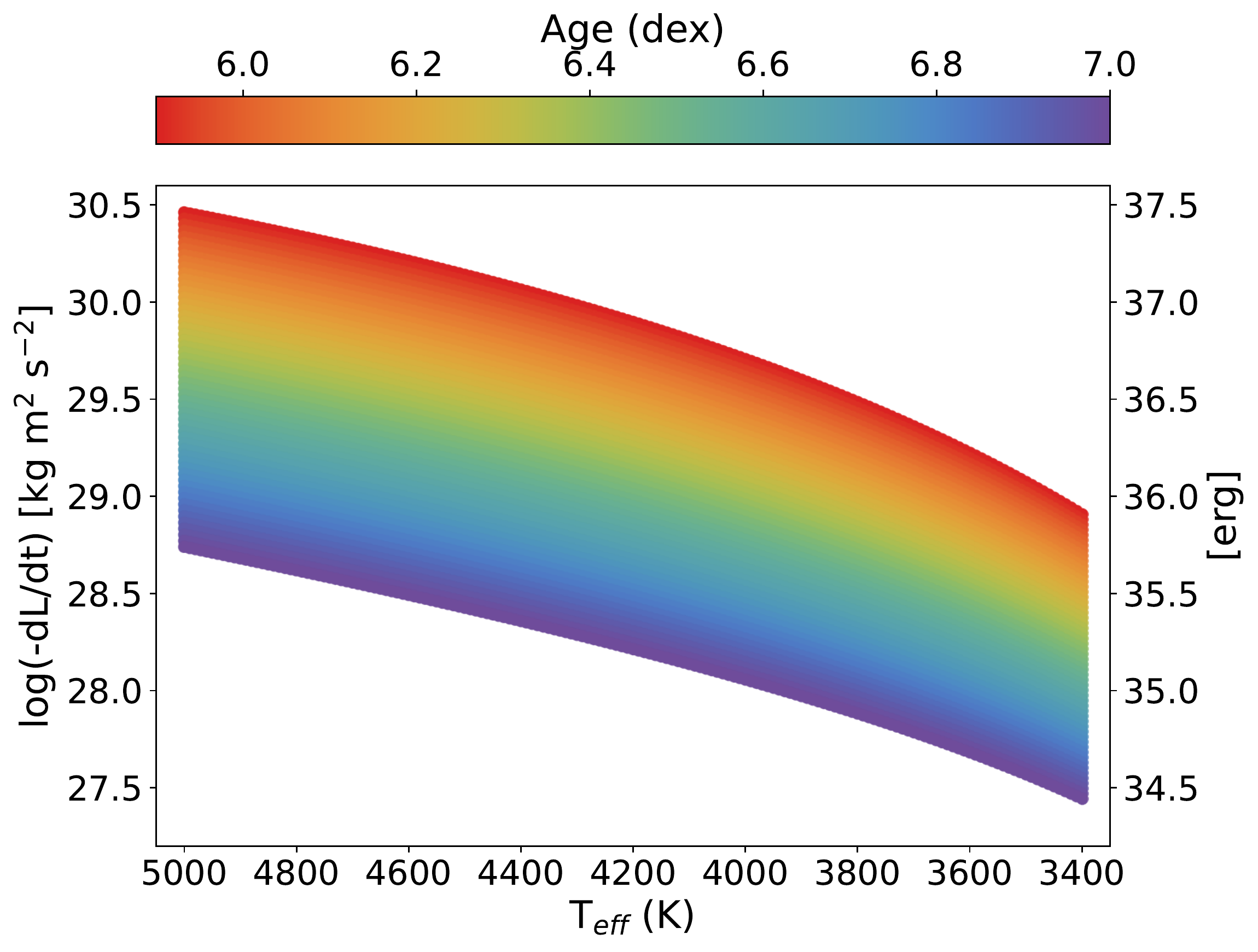}
\plotone{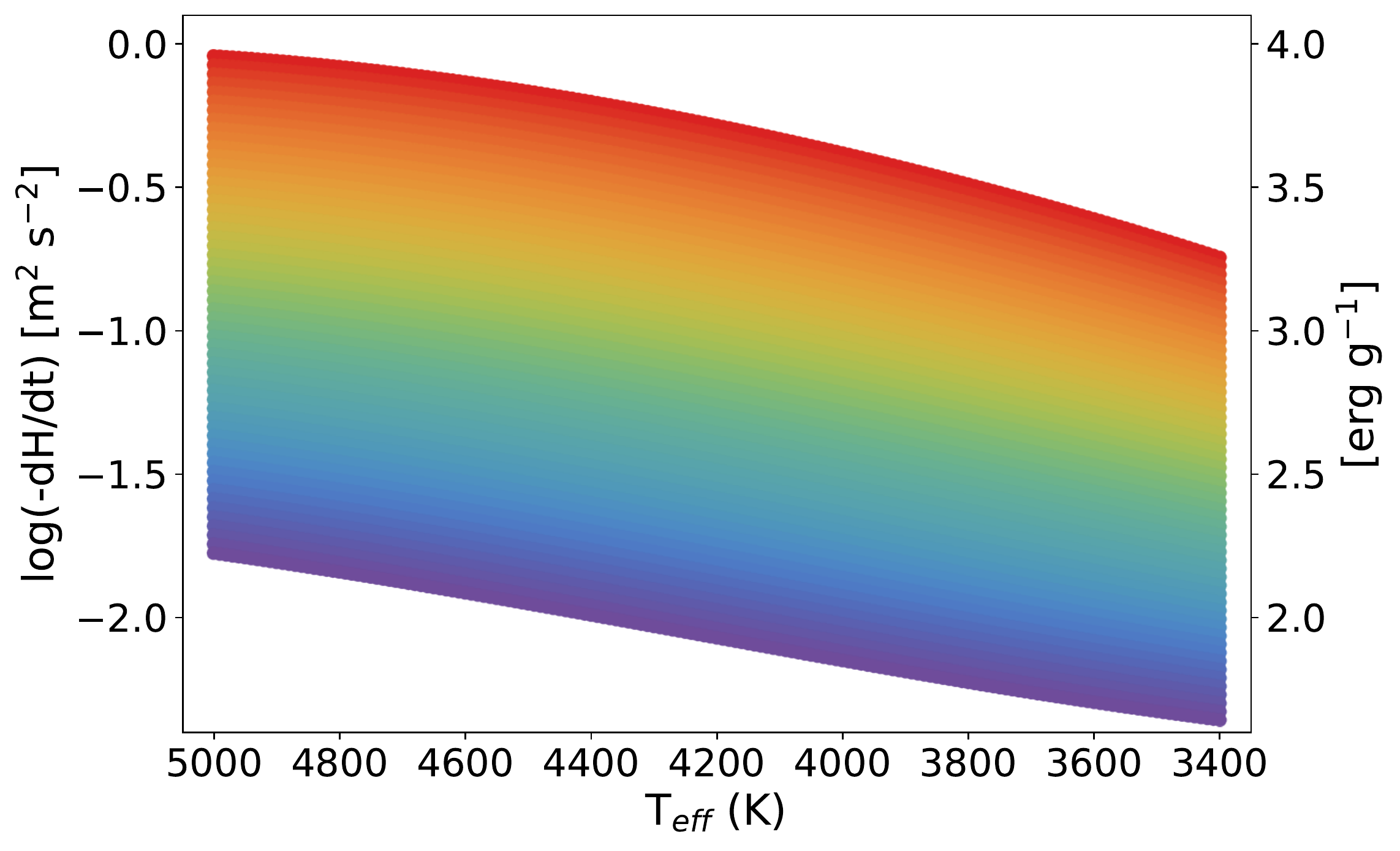}
\caption{Rate of loss of angular momentum $L$ (left) and specific angular momentum $H$ (right), as a function of temperature and age, assuming the empirical relation in Figure~\ref{fig:interp} and Table~\ref{tab:coeff}.
\label{fig:dl}}
\end{figure}

In the analysis of slow rotators, we limit the sample only to the sources with rotational periods $>$2 days. We examine the trends in this sample as a function of age (Figure \ref{fig:orion_sangmom}).

There is no appreciable difference in the distribution of periods relative to the ages of the stars. However, over the first 10 Myr, the radii of the stars rapidly contract. Because of this, there is a significant evolution in the angular momentum. We estimate $L$ assuming a solid body rotation via $L = 2/5MR^2\Omega$, where $\Omega=2\pi/P$. $L$ is not conserved during this contraction, as such the stars are losing angular momentum, most likely due to magnetic braking, despite the lack of significant evolution in rotational periods.

One complexity that can skew this approximation of $L$ is the differential rotation in stars, with rotation at the poles being slower than near the equator. At the moment we cannot determine the precise latitude of the spots. However, pre-main sequence stars moving on Hayashi tracks are expected to have significantly smaller differential rotation than, e.g., the Sun, with the shear of only 3.9\%, as it decreases with increasing convective zone depth \citep{kuker2001,marsden2011}. As such, a solid body approximation is reasonable for the young stars in this sample.

The resulting distribution of L appears to be mostly consistent between slowly rotating CTTSs and WTTSs, suggesting that the overall angular momentum content of the stars evolves largely irrespective of their disks. 
And, indeed, this spin down will continue throughout the lifetime of low mass stars, long past their shedding of the disks or entering the main sequence \citepalias{kounkel2022a}.

We bin the data using a kernel of $\pm$0.1 dex in age, $\pm$200 K in \teff, and find the running median in $L$ in each bin. Similarly to \citetalias{kounkel2022a}, using the resulting average $L$ in the 2D plane, we perform an empirical fit using a formalism of
\begin{equation}\label{eqn1}
\begin{split}
\log L=a_0+a_1\log T_{\rm eff}+a_2(\log T_{\rm eff})^2+a_3(\log T_{\rm eff})^3+\\
b_0\log t +b_1\log t\log T_{\rm eff}+b_2\log t(\log T_{\rm eff})^2
\end{split}
\end{equation}
\noindent where $\log T_{eff}$ is the $\log_{10}$ of \teff\ in K, and $\log t$ is $\log_{10}$ of age in years. The best fit is shown in Figure \ref{fig:interp}, and the coefficients are included in Table \ref{tab:coeff}. Independently, we also fit specific angular momentum per unit mass ($H$). We note that this functional form is arbitrary, and a comparable fit can be achieved with a different power dependence on both $T_{\rm eff}$ and $\log t$. In \citetalias{kounkel2022a} this form was chosen as the lowest power polynomial that could fit the age range from 10 Myr to 1 Gyr and be easily reversible with respect to age. In the Orion sample exending only up to 10 Myr, a maximum power of 2 instead of 3 can also result in a good fit, but for the sake of consistency we retain the same formalism.

\begin{deluxetable}{ccc}
\tablecaption{Fitted coefficients for gyrochronology relations
\label{tab:coeff}}
\tabletypesize{\footnotesize}
\tablewidth{\linewidth}
\tablehead{
 \colhead{Coefficient} &
 \colhead{Value ($L$)} &
 \colhead{Value ($H$)}
 }
\startdata
$a_0$ & $-$1926.1822 & 1554.6763 \\
$a_1$ & 1352.6072 & $-$1590.9434 \\
$a_2$ & $-$299.8859 & 524.9803 \\
$a_3$ & 21.2112 & $-$56.0318 \\
$b_0$ & 139.5928 & 186.4053 \\
$b_1$ & $-$76.1140 & $-$102.8078 \\
$b_2$ & 10.3335 & 14.1278 \\
\enddata
\end{deluxetable}

Through taking a partial derivative with respect to time of the resulting function, we observe the evolution in the rate of spin down (Figure \ref{fig:dl}). The loss of angular momentum is most effective at the youngest ages, with d$L$/d$t$ decreasing by more than an order of magnitude over the first 10 Myr. The primary source of angular momentum loss is likely to be magnetic breaking: young stars may have other factors that would affect their rotation (including accretion and outflows), but the combined net d$L$/d$t$ from these processes appears to be significantly smaller than the one we observe \citep[e.g.,][]{gallet2019}.

The spin down is more effective among the higher mass stars, this appears to be the case both with d$L$/d$t$ as well as d$H$/d$t$ (See also Section \ref{sec:empir} for discussion). The rate of loss of specific angular momentum appears to be somewhat proportional to $H$ itself; i.e., at a given age, a 5000~K star appears to have (on average) a specific angular momentum $\sim$3.6 times higher than a 3500~K star, and, similarly the ratio of d$H$/d$t$ for the same stars is also $\sim$3.6.

While the underlying trend of decreasing $L$ and $H$ at older ages is significant, at all ages there is significant scatter in their distribution, to a degree that is unlikely to improve with more precise measurements of mass, radius, or age. As such, while after a few hundred Myr a population of a given age would be able to form a relatively narrow gyrochrone \citep[e.g.,][]{douglas2019}, early on there is a wide range of the initial conditions pertaining to the angular momentum in the system that are yet to be erased. 

Nonetheless, it is striking that the angular momenta of the stars evidently follow much simpler, more monotonic relationships with stellar mass and age than rotation periods alone (i.e., compare various panels in Figure~\ref{fig:orion_sangmom} to the upper right panel in Figure~\ref{fig:orion_sangmom}). 

\subsection{Inclination}\label{sec:inclination}

The stellar radii measurements derived from the SED fitting are available from \citet{kounkel2018a}. An alternative way to estimate the radius is through a combination of the rotational period and rotational velocity, however as the latter is only available as \vsini, the resulting radius estimate \rsini\ would also have inclination dependence. As such, comparing these two estimates enables a constraint on \sini\ \citep[e.g.,][]{jeffries2007a,jackson2010}.

Mathematically, the upper limit of \sini\ is 1. In practice the ratio between \rsini/$R$\ does occasionally exceed 1 by more than the formal errors suggest, pointing to a systematic calibration issue in some of the measurements. Here we analyze some of the consideration that can lead to systematic issues that a) result in unreasonably large \sini\ and b) create a distribution of orientations of stars that, due to observational biases, do not appear to be entirely random.

\subsubsection{\teff}

Young stars tend to be strongly magnetically active, with a very high spot fraction coverage. Because of this, it may be difficult to describe these stars with a single temperature. Depending on the wavelength that is used to fit \teff, it is possible to have a systematic difference of a few 100 K depending on the wavelength of the spectrum \citep[e.g.,][]{lodieu2018}, as some regions of the spectrum are more sensitive to the temperature of the photosphere ($T_{\rm phot}$) or the spot ($T_{\rm spot}$), as opposed to the true \teff. The estimate of radius from the SED fitting is directly dependent on \teff, obtained via $R\propto\sqrt{F/T^4_{\rm eff}}$, where $F$ is the total flux.

We note that as the radii reported in \citet{kounkel2018a} relied on temperatures that was systematically hotter than what is adopted here. Without the renormalization with updated \teff, this systematic bias propagated from \teff\ to radii would further exaggerate the number of sources with \sini$>$1. As such, this demonstrates the improvement in temperature calibration in APOGEE Net compared to earlier estimates, significantly reducing the bias in \sini\ from \teff. However, without performing an explicit two temperature fit, spectroscopically measured temperature will be lower than the temperature of the unspotted photosphere, but higher than the actual \teff. Due to the resulting luminosity differences of templates, during the SED fitting $R$ can becomes underestimated.

For example, a star with $T_{\rm phot}=4000$ K and $T_{\rm spot}=$3500 K with 50\% spot coverage would have \teff\ of 3775 K. However, in the H band, the resulting spectrum would be best matched by a 3900 K template, resulting in $R$ difference of 10\%.

\subsubsection{Differential rotation}

As previously mentioned, differential rotation in stars with large convective zones is expected to be relatively small \citep{kuker2001,marsden2011}, affecting young stars only on a few \% level. However, if a spot is found at higher latitudes, the rotation period at the equator would be somewhat faster than the one that is observed. This will result in overestimating \rsini, which will also overestimate \sini.

\subsubsection{Binaries}\label{sec:binbias}

The excess in \sini$>1$ distribution is most pronounced for sources that have previously been identified as SB2, reaching values up to 10. In such sources, \vsini\ tend to be overestimated, as the line broadening in the spectra is affected by the presence of both stars in the system, which the spectroscopic fit of \vsini\ did not take into the account. Similarly, there may be additional issues, such as mismatch in the source of origin between \vsini\ and the rotational period, as well as the excess flux from the companion in the SED fitting.

Unresolved binaries that are widely separated enough to have their orbital motion being negligible in \vsini, will also have issue in their $R$ estimate from the SED fitting. As $F$ will be contaminated by the second star in the system, $R$ will be artificially inflated, not representative of the radius of either the primary or the secondary. Since $R$ is too large, but \rsini\ in these wide binaries be more similar to systems with just a single star, the resulting \sini\ will be underestimated. When we compare the distribution of slow and fast rotators, fast rotators appear to favor significantly smaller \sini\ than the slow rotators.

\citetalias{kounkel2022a} have found that fast rotators generally occupy the binary sequence in a given population. The difference in the \sini\ distribution between slow and rapid rotators is consistent with originating from unresolved binaries. As such, to provide a more equitable distribution of \sini\ that is dependent on other stellar properties, in evaluating other biases we focus solely on slow rotators. 

\begin{figure}
\epsscale{1.2}
\plotone{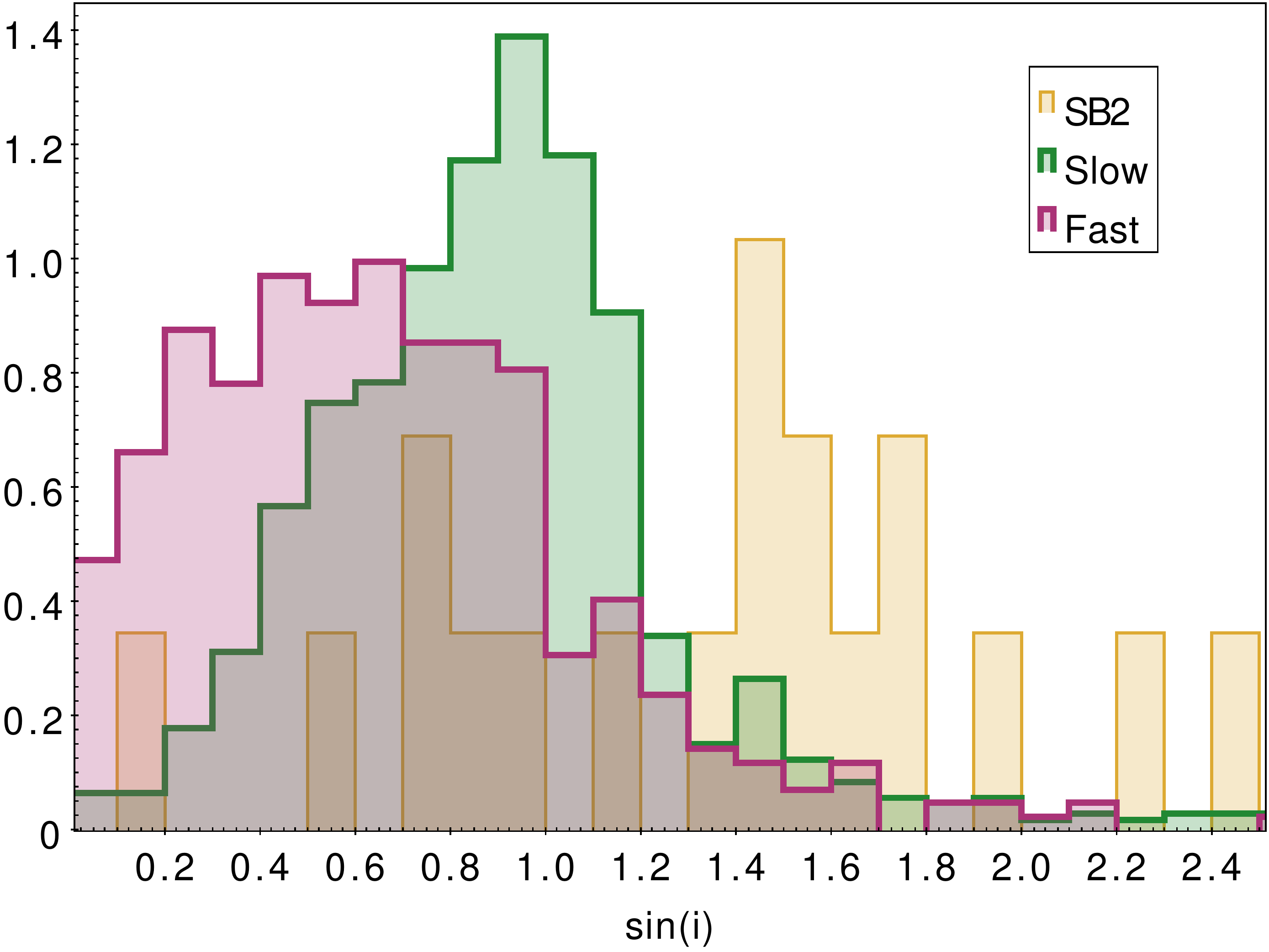}
\plotone{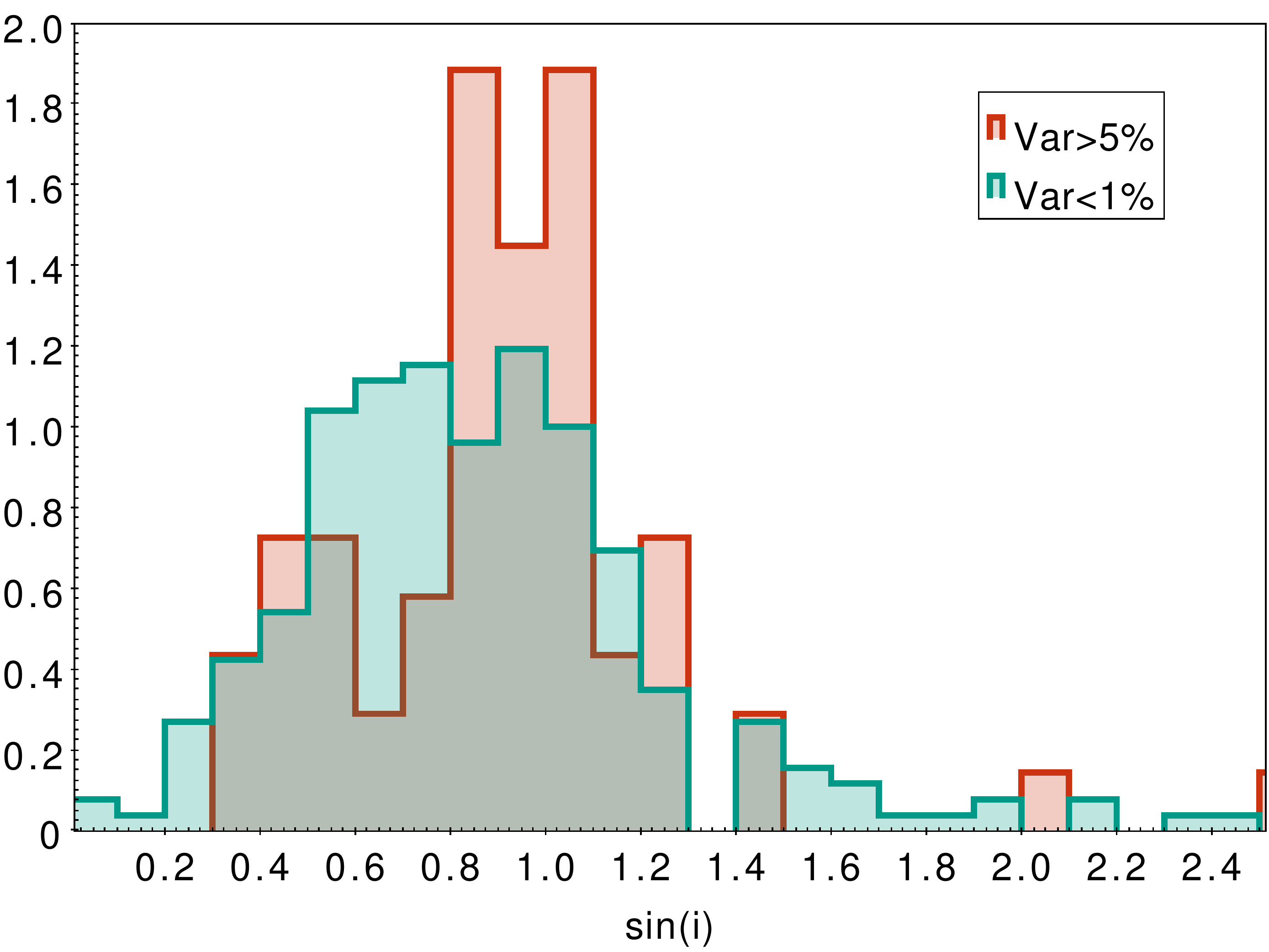}
\plotone{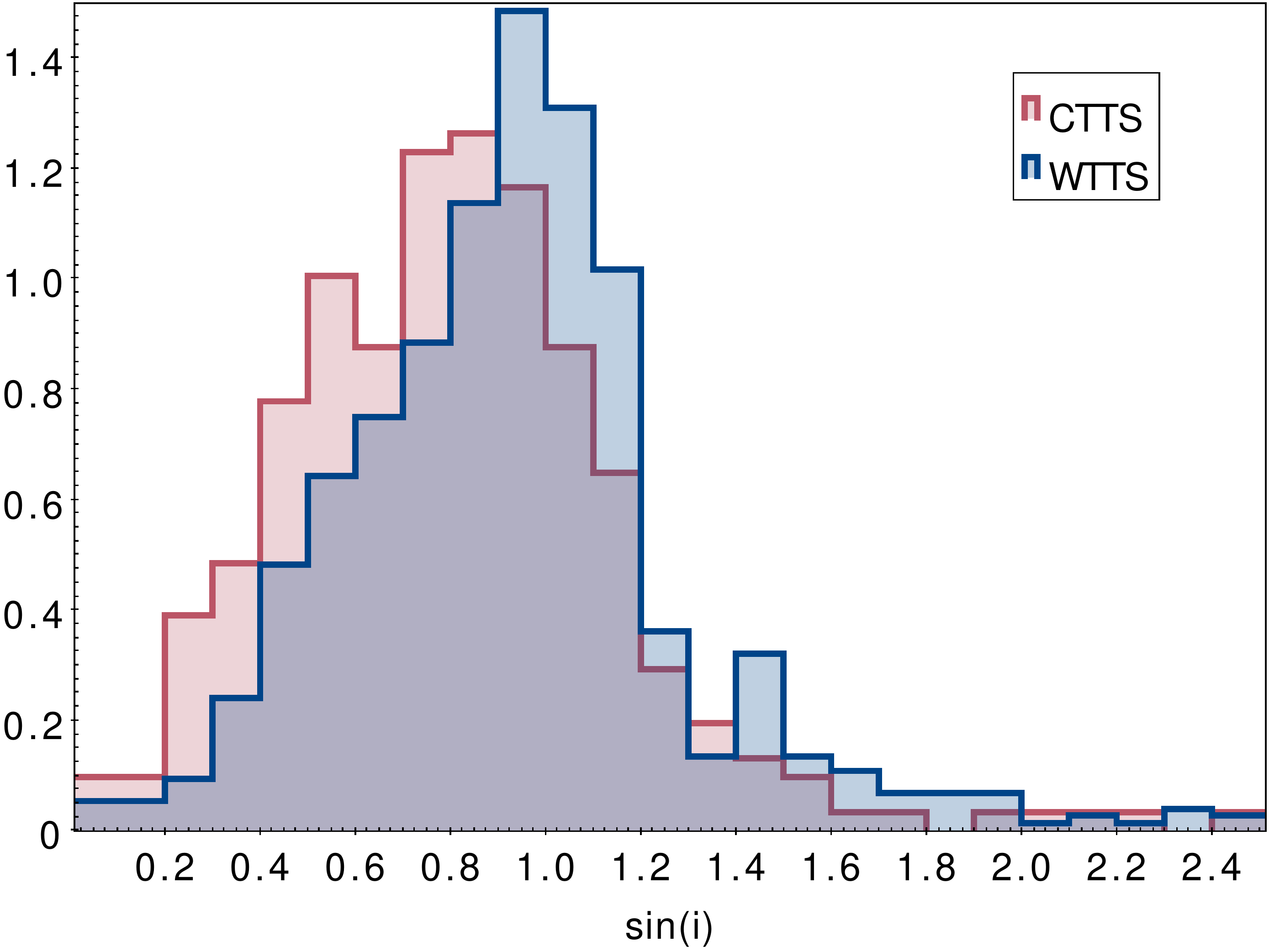}
\caption{Distribution of \rsini/R ratios, as a proxy of \sini. Top: a comparison between slow and fast rotators. Middle: a comparison of the sources with strong ($>$5\% of the total flux) and weak ($<1$\%) variability (limited just to the slow rotators with P$>$ 2 days), as well as double lined spectroscopic binaries from \citet{kounkel2019}. Bottom: a comparison of CTTSs and WTTSs, also limited just to slow rotators.\label{fig:sini}}
\end{figure}

\subsubsection{Detectability of periods and \vsini}

It is important to consider the biases of the underlying catalog that influence the resulting distribution. In order to be able to measure rotation period, a star (of any age) needs to be spotted, and these spots need to rotate in and out of view. The strongest signal is going to be produced by the edge-on systems. In the face-on systems, rotation cannot be observed. In the intermediate orientations, the systems with intrinsically larger spot sizes would be favored, since otherwise an induced variability from a weaker spot that may not fully rotate out of the field of view may be too shallow to be detected. We find that sources with a large amplitude of variability appear to favor edge-on inclinations, in comparison to other sources (Figure \ref{fig:sini}). If initially sources with strong and weak variability have spots of the same size (given that their ages and masses are comparable), weakly variable stars will have more intermediate inclinations, (resulting in a portion of the spot to never rotate out of the field of view or, alternatively, never rotate in the field of view), and strongly variable stars will be edge-on.

Similarly \vsini\ would also be biased against face-on systems, due to the necessary resolution in the spectra to confidently measure very weak broadening. As such, the expected distribution of inclination angles is not expected to be random, but rather favor \sini$\longrightarrow$1.

This does have implications for the stars in older populations in which magnetic activity is weaker, leading to smaller spots. At more advanced ages, if a star is detected as variable at more advanced ages, it is more likely to be edge on, as inclined systems might have the amplitude of variability less than the instrumental sensitivity.

\subsubsection{Extinction}

Young stars have additional biases in their detection. For example, \sini\ of CTTSs tends to be slightly smaller than \sini\ of WTTS stars. Applying a two-sided KS test shows that the two populations are different at 3$\sigma$ level. This difference is driven solely by the deficit of the edge-on CTTSs; restricting both samples to \sini$<$0.9, these inclined CTTSs and WTTSs are consistent to originating from the same distribution at $<$1$\sigma$ level.

CTTSs still have dusty disks, and an edge-on disk orientation would produce a highly reddened source, most likely rendering it to be too faint to be observed in a magnitude-limited survey like APOGEE. As such, while WTTSs would favor an edge-on orientation, CTTSs in the sample would tend to the sources where the photosphere is not fully obscured by the disk, but still as close to \sini$\longrightarrow$1 as possible within this constraint. We further discuss the implications of this in Section \ref{sec:agedep}.

\section{Discussion} \label{sec:disc}

\subsection{Empirical constraints on models of angular momentum loss}\label{sec:empir}

The fact that young, low-mass stars must somehow deplete their angular momentum by at least an order of magnitude prior to the main sequence has been understood as a challenge since the earliest measurements of their rotational characteristics \citep[e.g.,][]{cohen1979,Herbst:1982,Bouvier:1986,Basri:1987}. Whereas the rotational evolution of low-mass stars on the main sequence has long been well understood as the result of magnetized stellar winds acting on Gyr timescales \citep[e.g.,][]{skumanich1972,Kawaler:1988,barnes2001}, it has also been recognized that standard stellar winds cannot achieve the necessary angular momentum losses for PMS stars on the short timescale of $\sim$10$^7$~yr. Consequently, a number of mechanisms have been proposed that are unique to the PMS phase, most of which involve harnessing the inertia of circumstellar disks and/or the power of accretion from those disks during the disk lifetime of the first few Myr. 

For example, early attempts to model the exchange of angular momentum from star to disk in CTTSs were based on the model of \citet{Ghosh:1979}, in which the star transfers angular momentum to the inner part of the disk where the stellar magnetosphere intersects (and truncates) it. \citet{Shu:1994} developed a modified form of the \citet{Ghosh:1979} model---the so-called ``X-wind" model---that invokes the pinching of field lines in the immediate vicinity of the co-rotation radius in the disk to regulate the star's rotation, keeping the star's rotation period constant even as the star contracts (i.e., ``disk-locking"), and the excess stellar angular momentum ejected in a disk wind outflow \citep[e.g.,][]{Koenigl:1991}. 

Alternatives to the X-wind model include those that still utilize the star-disk interaction but remove the angular momentum from the star more directly. For example, the magnetospheric ejection model \citep[e.g.,][]{Zanni:2013} is based on MHD simulations of the sporadic opening and reconnection of magnetospheric field lines, which eruptively launch the magnetically confined mass; this is conceptually similar to stellar coronal mass ejections but with the long lever arm of the star-disk interaction region. Another example is the so-called accretion powered stellar wind model \citep[e.g.,][]{matt2012}, in which disk accretion itself acts to open stellar field lines and launch a powerful stellar wind, thereby extracting angular momentum directly from the stellar surface. 

All of these mechanisms make predictions for the torque that can be exerted on the star. With our empirically determined angular momentum loss rates for TTSs, anchored in directly measured rotation rates, distances, and empirically determined radii for a very large number of stars with ages 10$^6$--10$^7$~yr (i.e., Figure~\ref{fig:dl}), we are in a position to quantitatively confront the model predictions beyond previous comparisons to stellar rotation period distributions \citep[e.g.,][]{stassun1999,Rebull:2004,irwin2009}. 

The recent work of \citet{gallet2019} provides an especially useful basis for comparison of the model predictions to our measurements, as those authors constructed a family of models that attempt to self-consistently incorporate the action of all of the disk/accretion-based mechanisms described above (cf., their Figure~6). The models suggest that, for accretion rates that are not too high (i.e., $\dot{M} \lesssim 10^{-8}$ \msun/yr; otherwise, the net torque on the star is positive), spin-down torques as strong as $10^{36}$~erg are possible during the first $\sim$1~Myr and declining below $10^{35}$~erg after 1--2~Myr. By comparison, our measurements (Figure~\ref{fig:dl}) imply that the stars in fact experience torques as strong as $10^{37}$~erg during the first $\sim$1~Myr, declining to $\sim$10$^{36}$~erg after $\sim$2~Myr, and finally below $\sim$10$^{35}$~erg at $\sim$10~Myr. 
Thus, it appears that the disk/accretion-based mechanisms may not supply the full torque budget that we have measured, at least not on their own; an additional source of torque may be required that does not rely on disk inertia or accretion power.

One possibility is extreme coronal mass ejections (CMEs). 
\citet{Aarnio:2012} quantitatively explored this possibility by translating the empirical solar relationship between X-ray energy and CME ejected mass rates \citep[see][]{Aarnio:2011} to the observation by the Chandra Orion Ultradeep Project of extremely large flaring loops on TTSs in Orion \citep{Favata:2005}. \citet{Aarnio:2012} found that the stars can eject very massive CMEs with a high frequency during the first $\sim$10~Myr, and that the high observed CME losses are independent of whether the stars have retained their disks or are actively accreting \citep[see also][]{Aarnio:2013}. These studies estimated the torque from extreme CMEs as ranging from $10^{35}$~erg to $10^{37}$~erg \citep[cf., Figure 4 in][]{Aarnio:2012}, comparable to the angular momentum losses we have determined empirically in this work (Figure~\ref{fig:dl}).

\subsection{Nature of rapid rotators}\label{sec:raprot}

Previous observations have suggested that rapid rotators appear to be almost ubiquitously associated with unresolved binaries \citep{stauffer2016,stauffer2018,simonian2019,gillen2020a,bouma2021,kounkel2022a}. In populations with a clearly defined cluster sequence on the HR diagram, they tend to occupy the binary sequence Given the resolution of the data used to make this assessment, here ``unresolved'' binaries includes all of the separations up to several tens of au. 

In younger regions ($<$10 Myr) such as Orion, an age spread of just a few Myr is apparent in the HR diagram due to a rapid stellar evolution at these ages. As such, the binary sequence can be difficult to observe directly due to confusion with younger stars. However, as has been discussed in Section \ref{sec:binbias}, the \sini\ distribution of slow and fast rotators in Figure \ref{fig:sini} is consistent with the assumption of rapid rotators in Orion also being associated with wide but still unresolved multiples, similarly to what has been previously found older clusters.

We note that rapid rotators appear less likely to form around shorter period binaries with separation $<$ a few au. 
We examine their occurrence rate around SB1s \citep[which favors systems with periods $<$1000 days][]{kounkel2019}. There are 4 times fewer rapid rotators among SB1s than in the full sample (6\% vs 27\%). We discuss this in the following subsection. Thus, the range of separations of binaries hosting rapid rotators is expected to be a few to a few tens of au.

Given this range of separations, it may be possible to resolve these systems with high resolution imaging. A few such studies have been conducted in the Orion Nebula \citep{duchene2018,de-furio2019,de-furio2022} reaching separations down to 10 au, but, due to a limited sample size and the magnitude limit so far, there are no sources we identify as rapid rotators in common in these studies. In future, high resolution imaging of the sources identified as rapid rotators may be more fruitful in finding resolved binaries in comparison to a blind search.

\begin{figure}
\epsscale{1.15}
\plotone{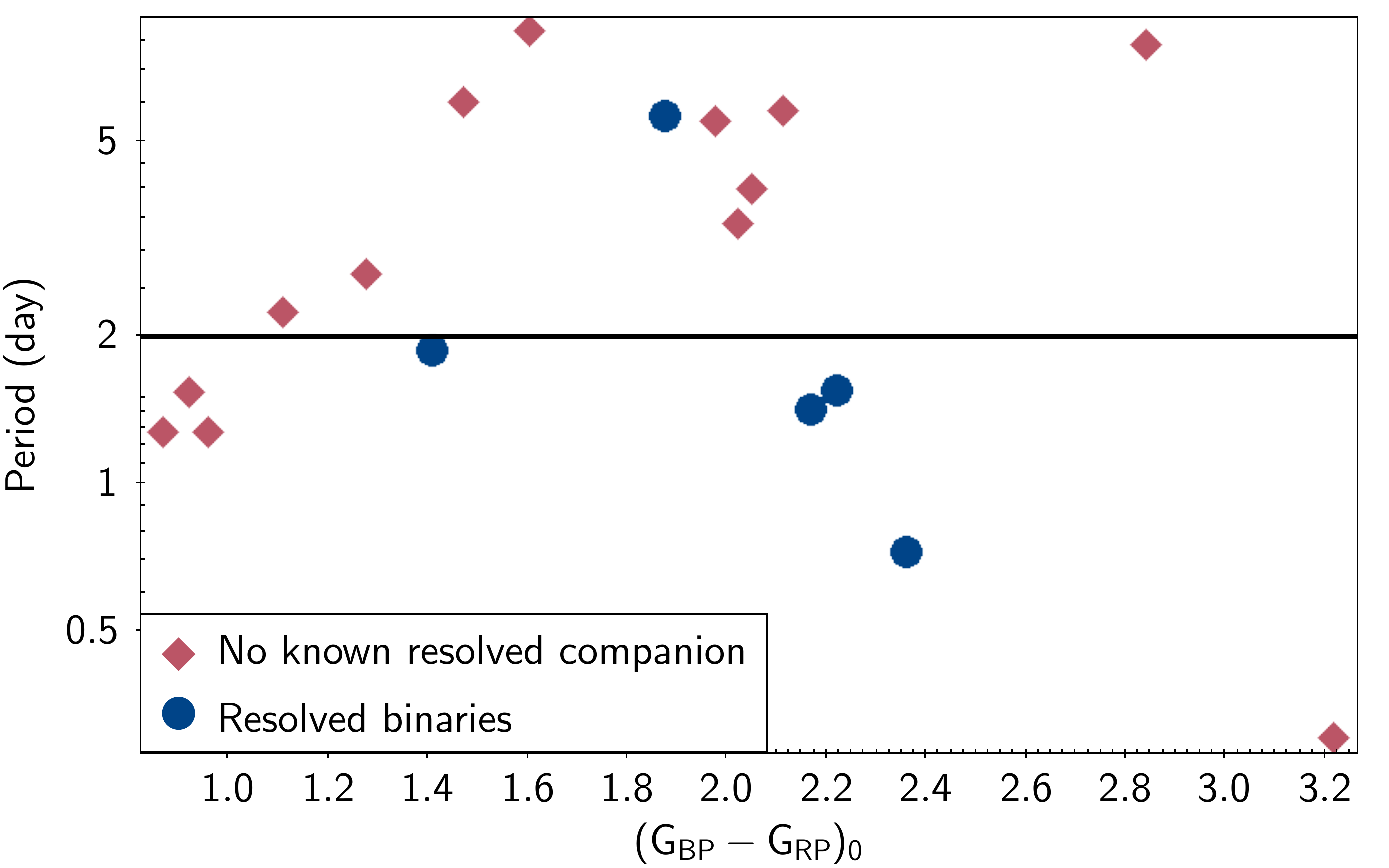}
\plotone{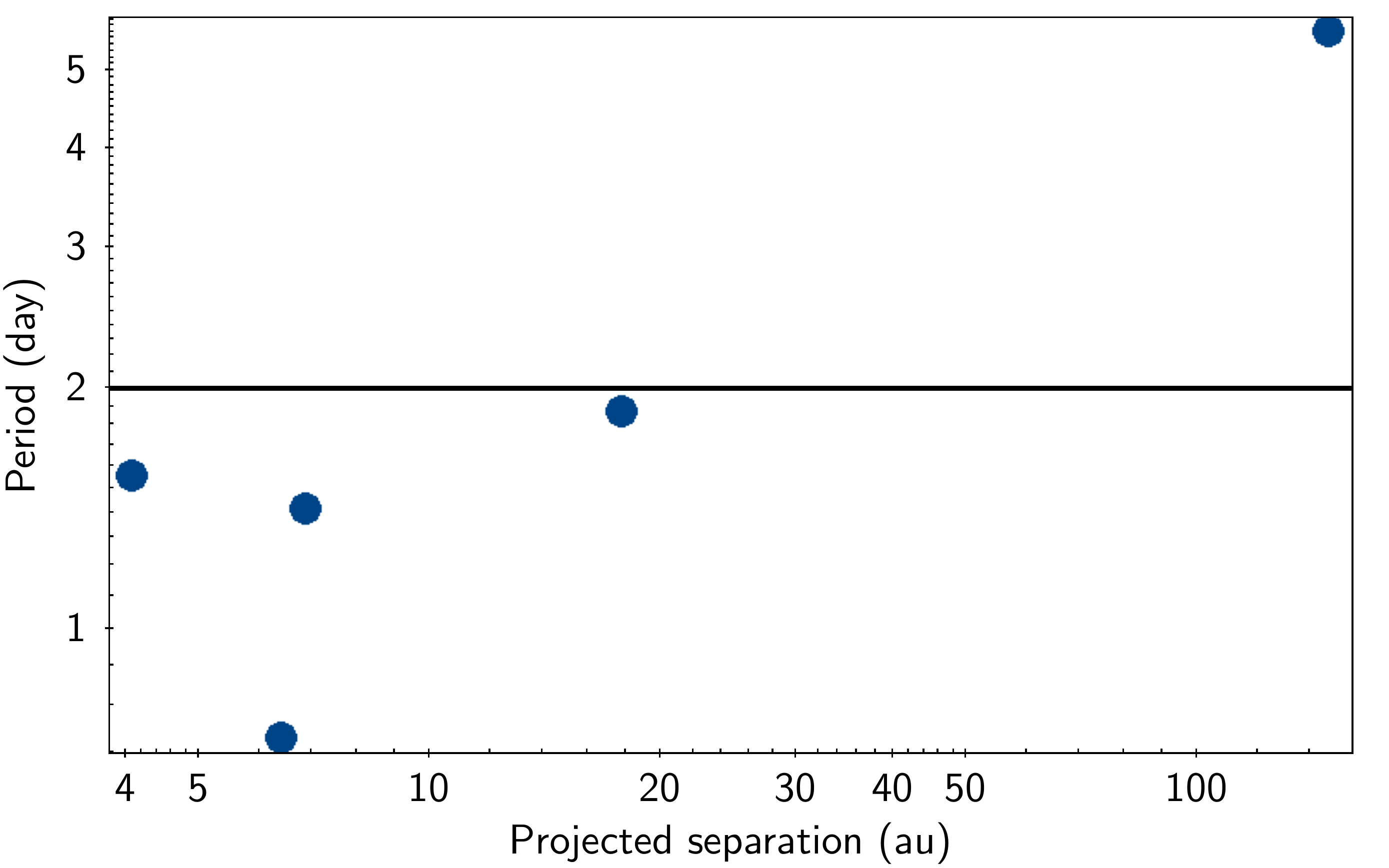}
\caption{Top: Rotational period as a function of color for the sample from \citet{kraus2011} in Taurus, with the periods from \citetalias{kounkel2022a}. Blue dots highlight binaries that have been resolved through high resolution imaging, red diamonds --- the sources in which no resolved companion has been identified to date. Black line shows the period of 2 days, delineating slow and rapid rotators \citep[see also Figure 4 in][]{stauffer2018}.  Bottom: rotational period as a function of projected separation in resolved binaries \label{fig:tau}}
\end{figure}

However, more in-depth studies have been conducted in other star forming regions, such as Taurus molecular clouds. Due to its proximity, resolved binaries have been detected down to separations of 3 au \citep{kraus2011}. Cross-matching this census with the rotational periods from \citetalias{kounkel2022a} yields a sample of 18 stars, of which 5 have been resolved as binaries. Four of these binaries are rapid rotators, with projected separations ranging from 4 to 20 au consistent with our interpretation (Figure \ref{fig:tau}). The sole non-rapidly rotating resolved binary has a separation of 150 au, which suggests that the systems with separations larger than few tens of au are less likely to result in spin-up.

\subsection{Disk evolution in rapid rotators}

We observed in Figure~\ref{fig:period} a clear connection between stellar rotation period and the presence of a disk, in the sense that the rapid rotators are overwhelmingly found among the WTTSs (i.e., diskless stars). This association of rapid rotation with the lack of a disk may be a direct consequence of the association between rapid rotation and binarity (see Section~\ref{sec:raprot}), and the fact that binaries with separations of a few au to tens of au tend to disperse their disks on faster timescales than single stars \citep[e.g.,][]{kraus2012}. 

While it is well understood that very close, tidally locked binaries (with orbital periods of only a few days) will naturally be forced to rotate rapidly \citep[e.g.,][]{Melo:2001}, it is not immediately intuitive why binary systems with separations up to several dozen au should imbue their stars with a greater amount of angular momentum. One possibility suggested by recent simulations of binary star formation \citep{kuruwita2017} is that binaries are less effective at removing angular momentum from the system: for example, in those simulations, a binary with separation of 45 au has an efficiency of only 42\% in transporting angular momentum away from the system via outflows, relative to a single star. 
In that case, our observation that the fastest rotators tend to be in binaries that are wider than the reach of tidal interaction would be a vestige of the binary formation process. 

\citet{kuruwita2017} also find that close binary systems with separations of 2.5 au have an efficiency of 87\% in transporting angular momentum away from the system in comparison to the single stars. That is to say, compact binaries are significantly less likely to have angular momentum excess in comparison to the wider systems with separations of a few tens of au. This is consistent with our observations of close binaries being less likely to develop rapid rotation. Most likely, this is due to the compact systems resembling single stars in their interaction with the protoplanetary disk, with the outflows being magnetocentrifugally driven, compared to the outflows being driven by magnetic pressure gradient in wider binaries \citep{kuruwita2017}.


It is interesting in this context that we have observed a small fraction of rapid rotators among CTTSs, and that this prevalence of rapid rotators among the disked stars appears to increase with time (see Figure~\ref{fig:rapid}). This has also been observed by \citet{serna2021}. We may again interpret these rapid rotators as likely being binaries as well, but whose 
orbital configurations permitted the disk to survive for some time (thus remaining as CTTSs for this duration). 
%
In that case, the increasing prevalence of rapid rotators among the CTTSs with time may suggest that the increased angular momentum deposited into the inner system by the binary formation process (see above) was partially stored in the disk, and then gradually accreted onto the stars over the disk lifetime, thus allowing the stars to later emerge as rapid rotators just like their diskless binary cousins.

\subsection{Age-dependent inclination bias} \label{sec:agedep}

\begin{figure}
\epsscale{1.15}
\plotone{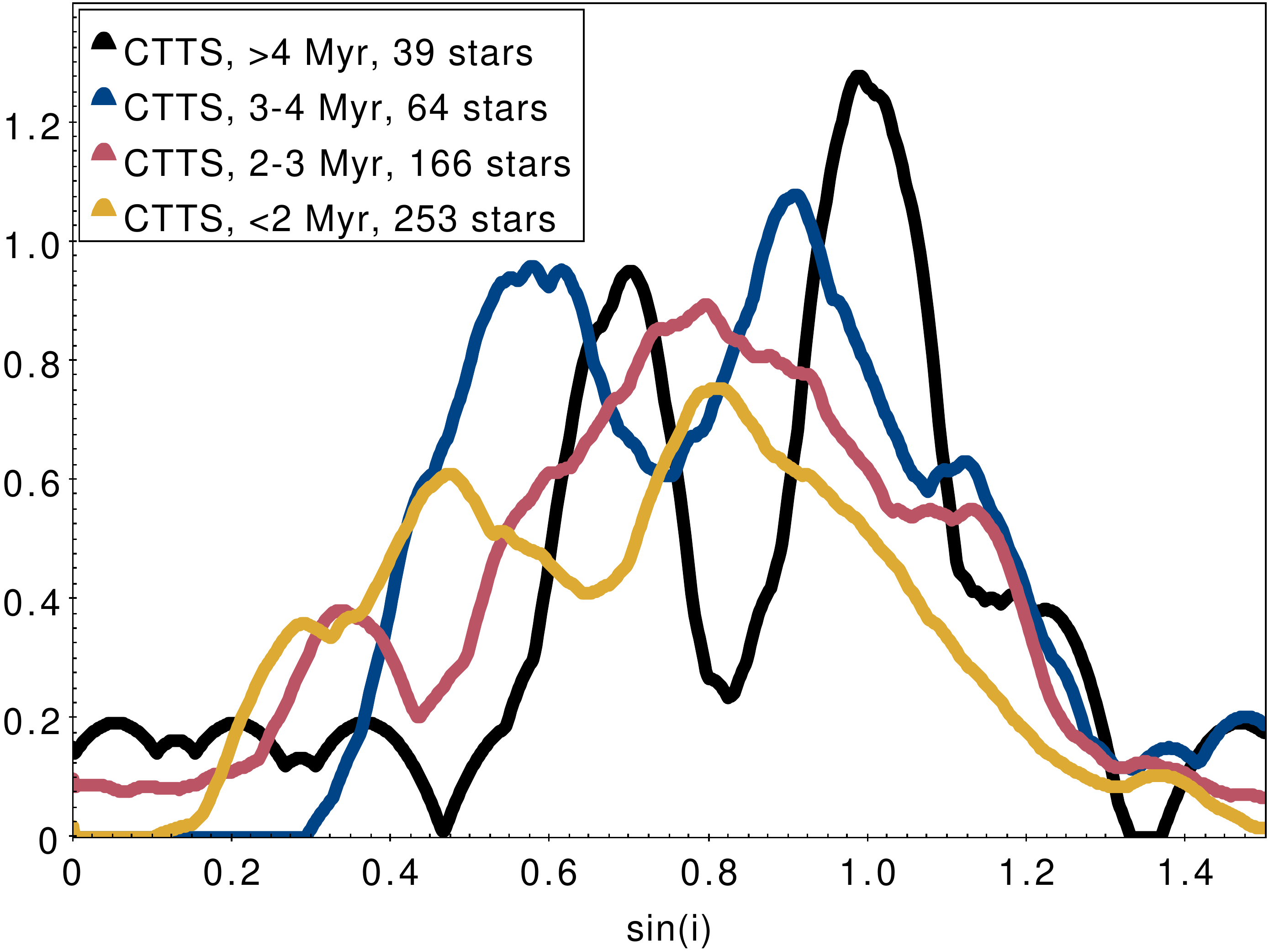}
\plotone{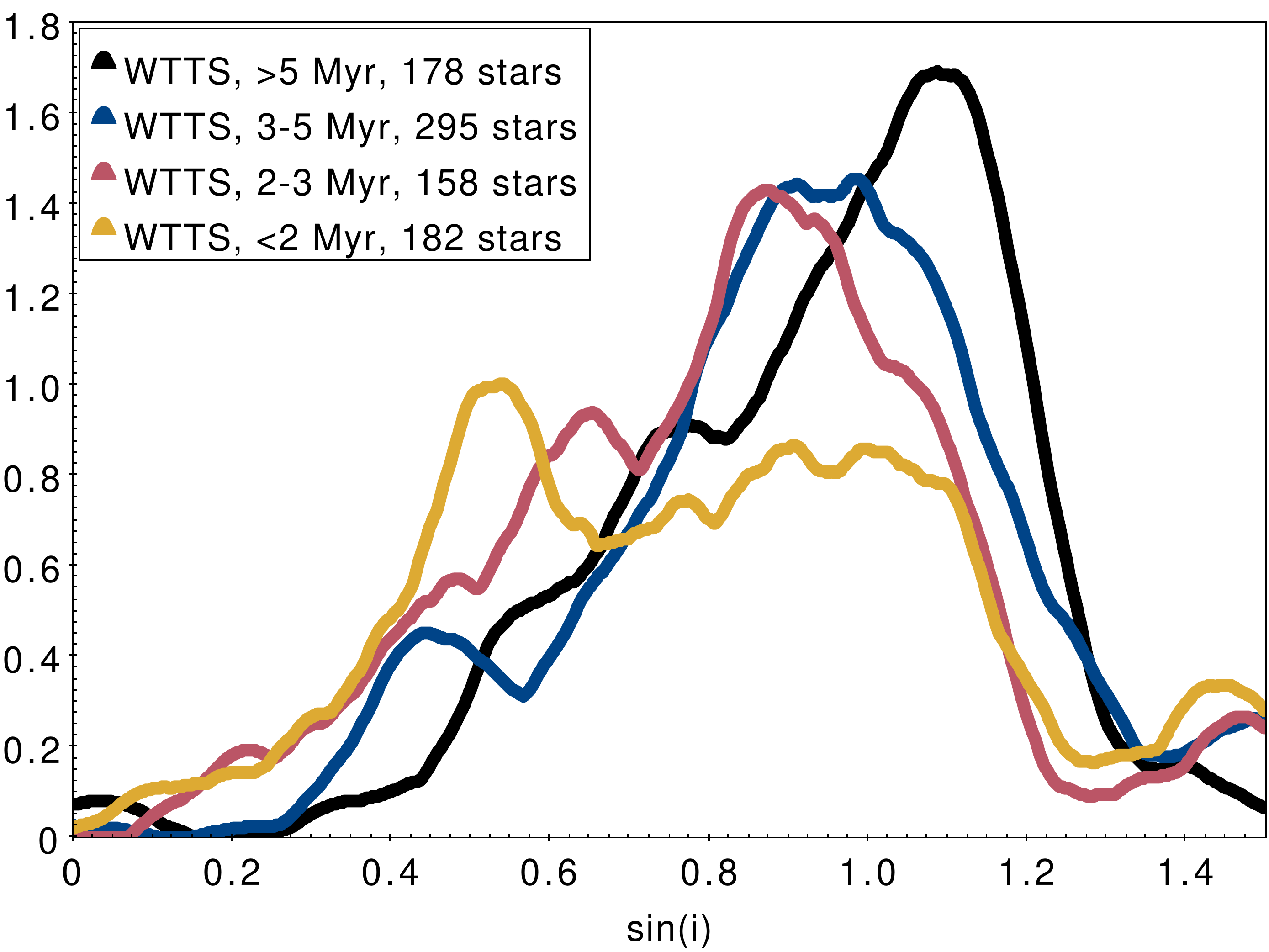}
\caption{Kernel density estimate distribution of \rsini/$R$ ratios of CTTSs and WTTSs as a function of age. Note that the last two age bins are different between the two panels, due to a decreased number of CTTSs at older ages. The distribution is often bimodal, the sources that appear to be inclined are likely to be unresolved binaries. \label{fig:sini1}}
\end{figure}

As can be seen in Figure \ref{fig:sini}, the average difference in \sini\ between CTTSs and WTTSs is $\sim$10--15\%. If we do assume that we see WTTSs across all inclinations between 0 and 90$^\circ$, then the maximum typical inclination of CTTSs in the sample would range between 0$^\circ$ and 55--65$^\circ$, with other inclinations being too extinct. This translates to the dust being found at the disk scale height of $\sim$0.5--0.6. This is in excess of what is typically found in protoplanetary disks \citep[scale height of $\sim$0.2 has been reported in some systems, but this is also considered to be quite high][]{natta2001,olofsson2013,montesinos2021}. Some dust can be present at $>3$ times the scale height, and some extinction can be expected even from the modestly inclined disks \citep{dalessio2006}.

However, the difference in the observed inclination may not necessarily be tied to the presence of the protoplanetary disk itself, but rather, the relative difference in age, owning to CTTSs being on average younger than WTTSs. If we compare \sini\ distribution for CTTSs and WTTSs separately across different age bins (Figure \ref{fig:sini1}), then the younger sources have a distribution of \sini\ that favors smaller values than those that are more evolved. This appears to hold true regardless of if the star has a significant IR excess from disk/has active accretion, or not. CTTSs still favor a slightly smaller \sini\ than WTTSs at a given age bin, but the difference is less drastic than in the overall distribution.

As such the source of opacity that reddens these stars likely comes from the outer debris disk/envelope that would not necessarily show significant IR excess in all but the longest of wavelengths, which may have been missed by the selection of disk-bearing stars. It is likely to settle or dissipate in a few Myr, but a puffy debris disk appears to be relatively common at ages $<$2 Myr, even if a star has already lost its inner disk. Regardless of the dominant source of opacity, be it the disk or the envelope, it is the dust along the line of sight that results in a deficit of young edge-on systems in a magnitude-limited sample. The younger a star is, the less likely is the dust to have been settled. No other age-dependent properties are likely to be responsible for this bias.

The difference in the distribution of inclinations between the stars of different ages may be an important challenge in being able to determine stellar ages. If the spots are not distributed randomly on the photosphere, but, e.g., spots are preferentially located along the pole instead of the equator, as has been demonstrated in several young systems \citep{donati2013, donati2014,donati2015} or more evolved stars with strong magnetic fields \citep[e.g.,][]{roettenbacher2016}, then this may impose a bias on the stellar age. The systems where the spots are prominently visible would appear to be less luminous than their unspotted counterparts. Coupled with the biases on the inclination, as well as the relative spot size and the spot contrast at different \teff\, this may be responsible for the trends in age as a function of mass in a given population in Figure \ref{fig:agecomp}.

When separated into different age bins, \sini\ distribution in Figure \ref{fig:sini1} often appears as bimodal. It is possible that the more inclined peak is dominated by the unresolved binaries. Although the rapid rotators are excluded from the sample presented in that figure, other binaries are still expected to be present in the sample, and, just like rapid rotators, they would have systematically underestimated \sini.

\section{Conclusions}\label{sec:conclusion}

Using TESS full-frame imaging data, APOGEE spectroscopic data, broadband spectral energy distributions, and Gaia parallaxes, we assemble the largest collection to date of stars younger than 10~Myr with rotation periods from \citetalias{kounkel2022a}, spectroscopic $T_{\rm eff}$ and $v\sin i$, disk classifications, empirically determined radii from \citet{kounkel2020}, and precise individual age estimates from \citet{mcbride2021}. With the resulting sample of $\sim$9000 stars in the Orion Complex, we examine the angular momentum content of young stars and the evolution of angular momentum with stellar age. 

Similarly to previous studies, we find that stars with rotation periods faster than 2~d are predominantly binaries, with typical separations of tens of AU. Such binaries are known to rapidly clear their disks, and consequently we observe an association between rapid rotation and disklessness. It is not clear how such binaries, which are much wider than the reach of tidal effects, come to possess greater amounts of total angular momentum, however it is consistent with recent simulations which find that outflows from binaries are less efficient at removing angular momentum from the inner system as compared to single stars. 

Among the (nominally single) stars with rotation periods slower than 2~d, we find that the angular momentum loss is most effective at the youngest ages, slowly decelerating its effect over time. Additionally, higher mass stars have higher total angular momentum (as well as specific angular momentum per unit mass) in comparison to their lower mass counterparts, but they experience more efficient angular momentum depletion, even at ages $<$10 Myr. 

More generally, we observe the familiar, gyrochronological horseshoe-shaped relation between rotation period and $T_{\rm eff}$, implying that the processes responsible for the universal evolution of stellar rotation on Gyr timescales are already in place within the first few Myr. We also find that our empirically quantified stellar angular momenta exhibit much simpler and monotonic relationships with stellar mass and age as compared to the relationships involving rotation period alone. We conclude that the relationship between rotation period and $T_{\rm eff}$ is largely a manifestation of the mass dependence of stellar structure and not of angular momentum per se. 


In addition, using the combination of rotational periods, rotational \vsini, as well as stellar radii measured from the SED fitting, we estimate the \sini\ of the stars, revealing a variety of biases in their distribution. We find that most variable systems tend to prefer an edge-on orientation, as this is most favorable configuration to observe star spots rotating in and out of view.
We also find that in a magnitude-limited sample, the younger systems tend to have on average lower \sini\ than the systems that are more evolved. This applies both to the stars with protoplanetary disks, as well as those that have already managed to disperse it. Most likely this difference in the distribution is driven by the extinction from the dust in the outer envelopes / debris disks, taking a few Myr to fully dissipate, even if the inner disk is fully gone. If there is a preferential distribution of spots along the photosphere, this difference in inclination distribution as a function of age may introduce a bias in the age estimate of some stars.

Most importantly, our directly measured rotation periods, together with precise Gaia distances and our empirically determined stellar radii for very large sample of stars with ages 1--10~Myr, permit us to construct empirical determinations of stellar angular momentum distributions and thus empirical determinations of angular momentum loss rates. Our measurements show quantitatively that the stars experience spin-down torques in the range $\sim10^{37}$~erg at $\sim$1~Myr to $\sim10^{35}$~erg at $\sim$10~Myr. Recent modeling efforts that utilize the inertia of circumstellar disks and/or the power of disk accretion to extract stellar angular momentum during the disk lifetime predict torques in the range $\sim10^{36}$~erg at $\sim$1~Myr to $\lesssim10^{35}$~erg by 2--3~Myr \citep[see, e.g.,][]{gallet2019}. Other mechanisms that do not depend solely on disk inertia or accretion power may therefore be required. For example, empirical observations of the extreme coronal mass ejections (CMEs) associated with X-ray super-flares observed in stars with ages 1--10~Myr appear capable of exerting torques ranging from $\sim10^{35}$~erg to $\sim10^{37}$~erg \citep[see, e.g.,][]{Aarnio:2012,Aarnio:2013}.
Our empirical measurements of the spin-down torques experienced by young stars should serve as a touchstone for these and other theoretical mechanisms of angular momentum loss in young stars.

\appendix
\restartappendixnumbering 

\section{Age consistency}\label{sec:age}

\begin{figure}
\epsscale{1.15}
\plotone{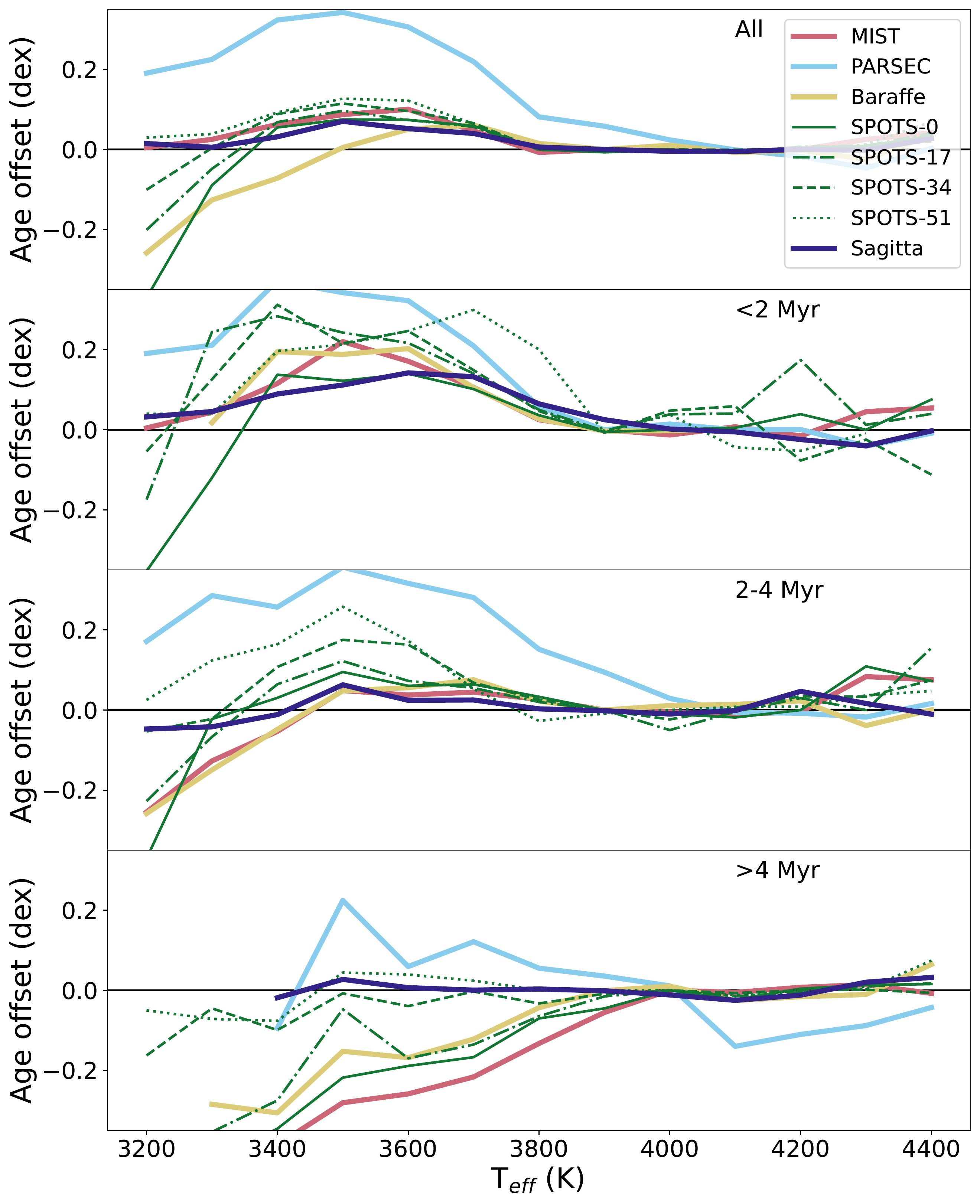}
\caption{Average systematic trends in the distribution of ages as a function of \teff\ across Orion, in different age ranges. Included models are MIST \citep{choi2016}, PARSEC \citep{marigo2017}, \citet{baraffe2015}, SPOTS with 0\%, 17\%, 34\%, and 51\% spot coverage \citep{somers2020}, and Sagitta \citep{mcbride2021}. The black line shows the constant age across all \teff.\label{fig:agecomp}}
\end{figure}

We estimate ages for stars through a variety of different means. Similarly to \citet{serna2021}, we have inferred the ages of these stars by using \teff, as well as Gaia and 2MASS photometry relative to various isochrones \citep{baraffe2015,choi2016,marigo2017,somers2020} using MassAge code (Hernandez J. in prep). We also use Sagitta \citep{mcbride2021}, which is a neural net trained on the photometry of young stars and the typical ages of star forming regions in which they are found. Finally, we also use spectroscopically derived \logg\ values, which strongly correlates with stellar age. All of these estimates are model-dependent. Although they all show similar progression of evolution, ordering sources from oldest to youngest in a similar manner, the relative zero point calibration, particularly as a function of mass, is systematically different between all of the approaches.

Previously, \citet{kounkel2018a} deconvolved the Orion Complex into 190 kinematically coherent subgroups. Collectively, these groups can be used to examine the spatial and temporal structure of the region. Independently, however, they are relatively compact and stars within a given group are expected to be coeval; i.e., they have self-consistent age distribution regardless of their mass.

\begin{deluxetable}{cc}
\tablecaption{Systematic offset in the mean age between different models
\label{tab:offset}}
\tabletypesize{\scriptsize}
\tablewidth{\linewidth}
\tablehead{
  \colhead{Model} &
  \colhead{$\log t_{\rm Sagitta}-\log t_{\rm Model}$ (dex)}
  }
\startdata
  MIST & +0.14\\
  PARSEC& +0.17\\
  Baraffe& $-$0.05\\
  SPOTS-0&+0.10\\
  SPOTS-17&+0.01\\
  SPOTS-34&$-$0.10\\
  SPOTS-51&$-$0.19
\enddata
\end{deluxetable}

Assuming this coevality, we can test the performance of various models in determining stellar ages, regardless of the absolute zero point calibration between them. For a given group, we estimate a median age that is reported by a given model, considering only the stars with \teff\ between 3800 and 4500 K. Then, we estimate an age in a small bin within 150 K of a given temperature, for all of the bins containing at least 4 stars, within each individual group. The difference between the two estimates shows the age gradient imposed by the underlying model as a function of mass. Afterwards, all of the groups of a given age range are averaged together to characterize the systematic trends.

We find that for \teff\ of 3800--5000 K, all of the models appear to be relatively stable, producing a coeval distribution of stars. However, this does not hold for cooler stars. In the populations younger than $<4$ Myr, isochrone models tend to overestimate the ages of stars by $\sim$0.1 dex, or $\sim$25\%, with the excess being more extreme for the youngest groups. PARSEC isochrones \citep{marigo2017} appear to have the worst performance, producing an excess of up to $\sim$0.3 dex, overestimating the ages of cool stars by a factor of $\sim$2. For groups older than 4 Myr, the trend is reversed, and the ages can be underestimated by as much as $\sim$0.2 dex. 

In all cases, strongly magnetic models with large spot coverage \citep{somers2020} tend to be offset from the same models without spots by $\sim$0.2 dex. But, because of the underlying trends in the data, in groups $<4$ Myr, non-magnetic tracks show greater coevality, whereas in groups $>$4 Myr, strongly magnetic tracks appear to be coeval. We note that magnetic tracks infer significantly older ages for a given group as a whole, with the difference of 0.3 dex between tracks with spot sizes of 51\% and those without spots (Table \ref{tab:offset}).

Across all ages, Sagitta \citep{mcbride2021} tends to have the greatest consistency in ages as a function of stellar mass, with only a modest overestimation in ages in groups $<$2 Myr (we discuss this trend in Section \ref{sec:disc}), but otherwise appears stable across other age ranges. In part, this is by construction, as the underlying model is empirical, trained on the average ages of stars in a given population across the solar neighborhood. The other models are theoretical, and may be lacking some of the physics for a precise representation of very cool stars.

For this reason, to ensure coevality between low mass and high mass stars in a given region, in subsequent sections we adopt ages from Sagitta. It's absolute calibration is most similar to the models from \citet{baraffe2015}, as well as the magnetic tracks with $\sim$17\% spot coverage. But, as the offset in age between the models is systematic, using a different set of models does not significantly influence the results presented in this work.

\software{TOPCAT \citep{topcat}, Plotly \citep{plotly}}

\acknowledgments

This work has made use of data from the European Space Agency (ESA)
mission {\it Gaia} (\url{https://www.cosmos.esa.int/gaia}), processed by
the {\it Gaia} Data Processing and Analysis Consortium (DPAC,
\url{https://www.cosmos.esa.int/web/gaia/dpac/consortium}). Funding
for the DPAC has been provided by national institutions, in particular
the institutions participating in the {\it Gaia} Multilateral Agreement.

\bibliographystyle{aasjournal.bst}
\bibliography{ms.bbl}

\end{document}